\documentclass[journal,review,submit,pdftex,moreauthors]{Definitions/mdpi} 
\firstpage{1} 
\pubvolume{1}
\issuenum{1}
\articlenumber{0}
\pubyear{2023}
\copyrightyear{2022}
\datereceived{} 
\daterevised{}
\dateaccepted{} 
\datepublished{} 
\hreflink{https://doi.org/} % If needed use \linebreak
\Title{The LHAASO PeVatron bright sky: what we learned}

\TitleCitation{The LHAASO Pevatron Era}

\Author{Martina Cardillo $^{1}$\orcidA{}* and Andrea Giuliani $^{2}$\orcidB{}}

\AuthorNames{Martina Cardillo and Andrea Giuliani}

\AuthorCitation{Cardillo, M.; Giuliani, A.}
\address{%
$^{1}$ \quad INAF -- Istituto di Astrofisica e Planetologia Spaziali,\\
  Via del Fosso del Cavaliere 100, 00133 Roma (RM), Italy\\
$^{2}$ \quad INAF -- Istituto di Astrofisica Spaziale e Fisica Cosmica,\\
  Via Alfonso Corti 12, 20133 Milano (MI), Italy\\
}

\corres{Correspondence: martina.cardillo@inaf.it}

\abstract{The recent detection of 12 \gray{} Galactic sources well above $E>100$ TeV  by the LHAASO observatory has been a breakthrough in the context of Cosmic Ray (CR) origin search. 
Although most of these sources are unidentified, they are often spatially correlated with leptonic accelerators, like pulsar and pulsar wind nebulae (PWNe). This dramatically affects the paradigm for which a \gray{} detection at $E>100$ TeV implies the presence of a hadronic accelerator of PeV particles (PeVatron). Moreover, the LHAASO results support the idea that sources other than the standard candidates, Supernova Remnants, can accelerate Galactic CRs. 
In this context, the good  angular resolution of future Cherenkov telescopes, such as the  ASTRI Mini-Array and CTA, and the higher sensitivity of future neutrino detectors, such as KM3NeT and IceCube-Gen2, will be of crucial importance.
In this brief review, we want to summarize the efforts done up to now, from both theoretical and experimental points of view, to fully understand the LHAASO results in the context of the CR acceleration issue.
}


\def \gray {$\gamma$-ray}
\def \grays {$\gamma$-rays}

\begin{document}

%%%%%%%%%%%%%%%%%%%%%%%%%%%%%%%%%%%%%%%%%%
\section{Introduction}
%%%%%%%%%%%%%%%%%%%%%%%%%%%%%%%%%%%%%%%%%%

\begin{figure}[!h]
    \centering
    \includegraphics[width=0.5\textwidth]{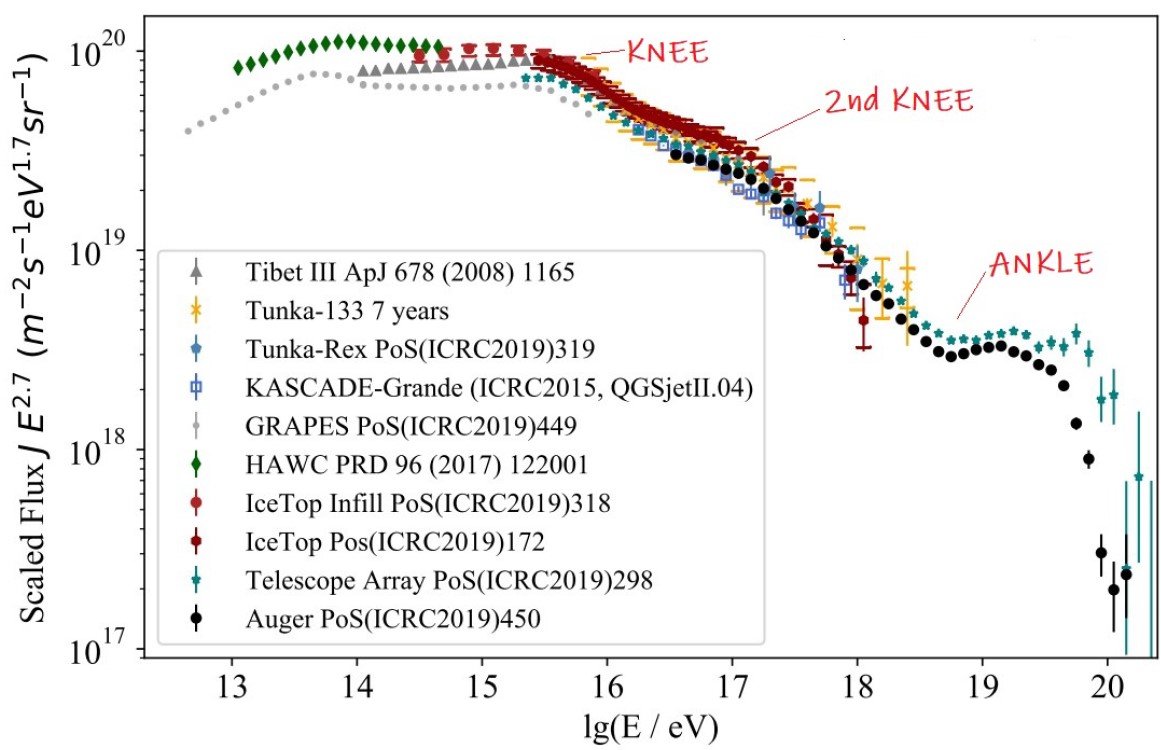}
    \includegraphics[width=0.43\textwidth]{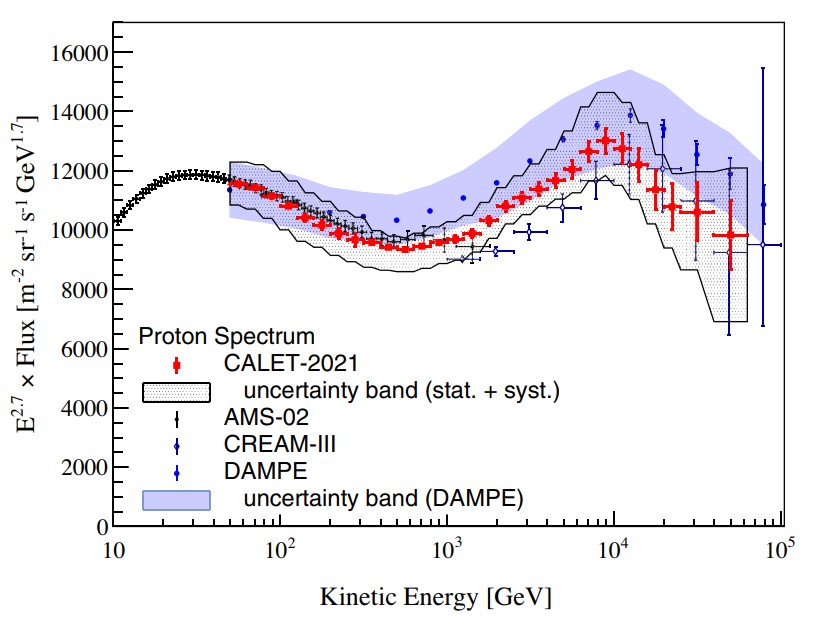}
    \caption{Left: Energy spectrum of HE CRs obtained from air-shower measurements (Figure from \cite{Schroder19_ICRC}). Right: Proton spectrum measured by CALET where the it is evident the not "pure" PL behavior of the CR spectrum before the knee (Figure from \cite{Adriani22_CALET}).}
    \label{Fig:CRs}
\end{figure}

The origin of Cosmic Rays (CRs) is among the most studied topics in High-Energy (HE) astrophysics.
CRs are relativistic particles, mainly protons ($92\%$) and ions, that fill our Galaxy. Their very extended spectrum spans from a few MeV to beyond $10^{20}$ eV (see Fig.\ref{Fig:CRs}, left). Historically, it was described by a single power-law with an index $\alpha=2.7$ up to the so called "knee" ($\sim 3\times10^{15}$ eV), due to Galactic CR contribution. However, the recent results of CALET/DAMPE show that the region below 1 PeV may not be as featureless as we thought, showing hardening and steepening between 100 GeV and 100 TeV \citep{Kobayashi21, Lipari20, Donghwa22H, Vink22H, Aharonian22H} (see Fig.\ref{Fig:CRs}, right).  
After the "knee", there is a spectral steepening ($\alpha=3.1$) up to the so-called "ankle" ($\sim 3\times10^{18}$ eV), where the spectrum hardens again, likely due to the emerging contribution of the extra-galactic component. A "second knee" at $\sim 10^{17}$ eV is probably due to heavy nuclei and, according to the last experimental results, the transition from Galactic to the extra-Galactic component is likely between the "second knee" and the "ankle" energies \citep{Schroder19_ICRC, Coleman19_secondKnee, Cristofari21}. 

The complexity of the CR spectrum perfectly fits in the HE Astrophysics context of these last years, especially after the last results of the LHAASO Observatory.

%!!!!!!!!!!!!!!!! 
%!!!!!!!!!!!!!!!! 

%---------------------------------------------
\subsection{Status of the field before the LHAASO results}
\label{Sec:B.L.}
%---------------------------------------------

One of the main channels to investigate the nature of Galactic CRs and their features is the non-thermal HE ($E>100$ MeV) \gray{} emission, produced either by electrons, mainly by Bremsstrahlung and Inverse Compton (IC) processes, and by protons via pion decay from p-p and p-$\gamma$ interactions. The discrimination between the two main types of processes, leptonic or hadronic, is at the base of CR acceleration phenomena. The \gray{}detection of the so called "pion-bump" at about 100 MeV (due to $\pi^{0}$ rest mass) is one of the direct proofs that confirm the presence of energized CRs in a source and until recently, a detection of Ultra High Energy (UHE, $E>100$ TeV\footnote{This energy threshold for the UHE definition is valid in the \gray{} context. If we consider particle context, the UHE range begins above $E>10^{17-18}$ eV}) \gray{} photons was considered unquestionable evidence of freshly accelerated CRs. Indeed, at these energies, electron IC should be limited by the Klein-Nishina regime\footnote{In the relativistic regime, $h\nu_{i}>m_{e}c^{2}$, the Compton Scattering is inelastic and characterized by $\sigma_{KN}$, derived by the Quantum ElectroDynamics, and not by $\sigma_{T}$} and UHE \gray{} should be explained only through the decay of neutral pions produced by collisions of PeV protons (CRs) with target protons \citep[for recent reviews see][]{Blasi19, Gabici19}. According to this view, the term "PeVatrons" has been coined to indicate Galactic CR accelerators (sources emitting \gray{} at E>100 TeV).

In the standard paradigm, Supernova Remnants (SNRs) are the main contributors to Galactic CRs \citep{Cristofari21}. We know that they energize CRs at the low \gray{} energies (MeV-GeV) thanks to their detection around the "pion-bump" energies by AGILE and Fermi-LAT \citep{Giuliani11, Ackermann13}, likely even through re-acceleration \cite{Cardillo16}, but in any SNR were detected \gray{} photons with E > 100 TeV. According to theoretical models, the reason is that SNRs can accelerate CRs at these energies only in the first 100 years of their life and all the known SNRs are older \citep{Bell13, Cardillo15}. The only way we can detect their UHE emission is in case CRs were trapped in near Giant Molecular Clouds (GMCs), emitting UHE photons even for $t>100$ yrs \citep{Gabici19, DeOnaWilhelmi20, Sushch23}. 

The lack of SNRs at UHE supports the chance that other sources can accelerate CRs and this possibility was largely investigated. In 2011, the Fermi-LAT satellite observed extended \gray{} emission associated with the superbubble surrounding the Cygnus OB2 region, then called "Cygnus Cocoon", confirming the hypothesis, proposed for the first time by \cite{Bykov01}, that star forming region are able to accelerate particles. Deeper analysis of this region, and further Very High Energy (VHE, 100 GeV$<E<$100 TeV) detection from other Massive Star Clusters (MSCs), supported this chance (see Section \ref{Sec:MSCs} and the Section on LHAASO J2032+4102 (\ref{Sec:Cocoon}) for more details). 

Later, the detection of several Pulsar Wind Nebulae (PWNe) at $E>56$ TeV by different instruments \citep{Abeysekara19_Crab, Abeysekara2020_56TeV, HESS19_J1825, Amenomori19_Tibet_Crab} and in particular of the Crab Nebula at $E>150$ TeV \citep{Abeysekara19_Crab, Amenomori19_Tibet_Crab} opened the possibility that also the PWNe, leptonic accelerators "by definition", could accelerate hadrons and, consequently, CRs. Moreover, UHE \gray{} detection from the Crab Nebula made evident that leptons are able to emit beyond the Klein-Nishina limit. Several works have shown that leptonic emission through IC can dominate at those energies in radiation dominated environments \citep{Breuhaus21} or in the surroundings of high spin-down powers pulsars \citep{Albert21, DeOnaWilhelmi22}.

In this new context, another fundamental channel comes into play: the neutrino detection. This is an unquestionable hint of CR acceleration, since neutrinos can be produced only by the decay of charged pions produced by p-p and p-$\gamma$ interactions \citep{Anchordoqui14}. Looking for neutrino detection in correspondence of PeVatron candidates is the main instrument to confirm the nature of \gray{} emitting sources \citep{Celli20}. These neutrinos might be detected by current (IceCube \citep{Abbasi09_IceCube} and Baikal-GVD \citep{Zaborov20_Baikal}) and future (P-ONE \citep{Agostini20_P-ONE}, IceCube-Gen2 \citep{Aartesn21_IceCube_Gen2}, KM3NeT-ARCA \citep{AdrianMartinez16-KM3}) VHE neutrino detectors \citep{Mandelartz15, Huang22_neutrini_b, Sarmah23_neutrini}. However, up to now, no neutrinos from Galactic sources have been detected with high significance \citep{Aartsen20_IceCube}. We have a slight evidence of neutrino emission only from the two extra-galactic sources: TXS 0506+056 (a blazar) and NGC 1068 (a Seyfert/starburst), associated with IceCube HE neutrinos at a significance around 3$\sigma$ \citep{Abbasi18, Aartsen20_IceCube}. Also most of the recent diffuse neutrinos detected over a period of 7.5 years \citep{Abbasi21_neutrini} are found to be of extra-galactic origin \citep{Chakraborty15, Petropoulou17, Sarmah22a}. 

%---------------------------------------------
\subsection{Status of the field after the LHAASO results}
\label{Sec:A.L.}
%---------------------------------------------

The recent LHAASO discovery of several Galactic sources emitting UHE \grays{} \citep{Cao21, Cao21_J0341, Aharonian21_J0621} represents a strong breakthrough in the context of PeVatron search. Most of the  sources seem to correlate to pulsars, and/or their nebulae, PWNe (e.g., the Crab Nebula for all), a class of well known leptonic accelerators.
These observations indicate the presence in these sites of ultra-relativistic electrons with energies reaching at least a few $10^{15}$ eV. As a consequence, the \gray{} detection at UHEs cannot be itself considered the proof of hadronic acceleration. 
This challenges the definition itself of “PeVatron”. In this work, we will call "PeVatron" generically an object capable of accelerating particles (hadrons or leptons) up to the PeV ($=10^{15}$ eV) range (and emitting \gray{} above 100 TeV). Consequently, to indicate a source that accelerates specifically hadrons (CRs), we need to use the term "hadronic PeVatron".

The LHAASO results also suggest that many classes of sources can accelerate CRs up to PeV energies: SNRs, MSCs, and maybe also the PWNe \citep[see][and references therein]{Vink22}. 
In the case of PWNe, there is also another complication: the standard first order Fermi acceleration mechanism, underpinning the Diffusive Shock Acceleration (DSA) theory, could be not the main one \citep[see][for a review]{Amato2021, Fiori22} but magnetic reconnection (MR) may be a valid alternative even if simulations suggest that it may be a process producing impulsive events (flares) but no a stable VHE emission \citep{Uzdensky11_Recon,Cerutti14_Recon, Lyutikov18_Recon}. This discussion, however, is beyond the aim of this review. 

Actually, some of the LHAASO PeV detection regions could be correlated with SNRs, keeping open the chance to have PeV emission from this kind of sources even after their 100 yrs \citep{MAGIC21_Boomerang, Liu22_Boomerang, Fang22_Boomerang, Liang22_Boomerang}. Several studies are in progress in order to establish a "look up table" of parameters useful for future instruments to distinguish which SNR is hadronic \citep{Corso23, Sudoh23}.

The association of VHE/UHE LHAASO emission with a determined kind of source in the observed region is challenging since the very low angular resolution of LHAASO and similar Extended Air Shower (EAS) arrays. For this reasons, future Imaging Atmospheric Cherenkov Telescopes (IACTs) as ASTRI Mini-Array \citep{Pareschi13, Scuderi19, Giro19, Scuderi22, Vercellone22} and CTA \citep{CTA19} will make the difference (see Fig. \ref{Fig:performance_curves}, right) (see Section \ref{Sec:future}).

The LHAASO results triggered several studies on the constraints given by potential neutrino emission on the \gray{} fluxes from the candidate PeVatron sources. In \cite{Huang22_neutrini_a}, the authors exploit the IceCube measurements in the surrounding of the LHAASO sources in order to constrain their possible hadronic contribution and in the following paper \cite{Huang22_neutrini_b} the same authors carried on a Bayesian analysis on the 10-years IceCube data (combined with the ANTARES \citep{Albert17_ANTARES} ones) with the same aim. The results obtained with the two approaches are very similar and can be summarized with the impossibility of strong constraints on the LHAASO sources (but the Crab) with the current instruments. 

\begin{figure}[!h]
    \centering
    \includegraphics[width=0.48\textwidth]{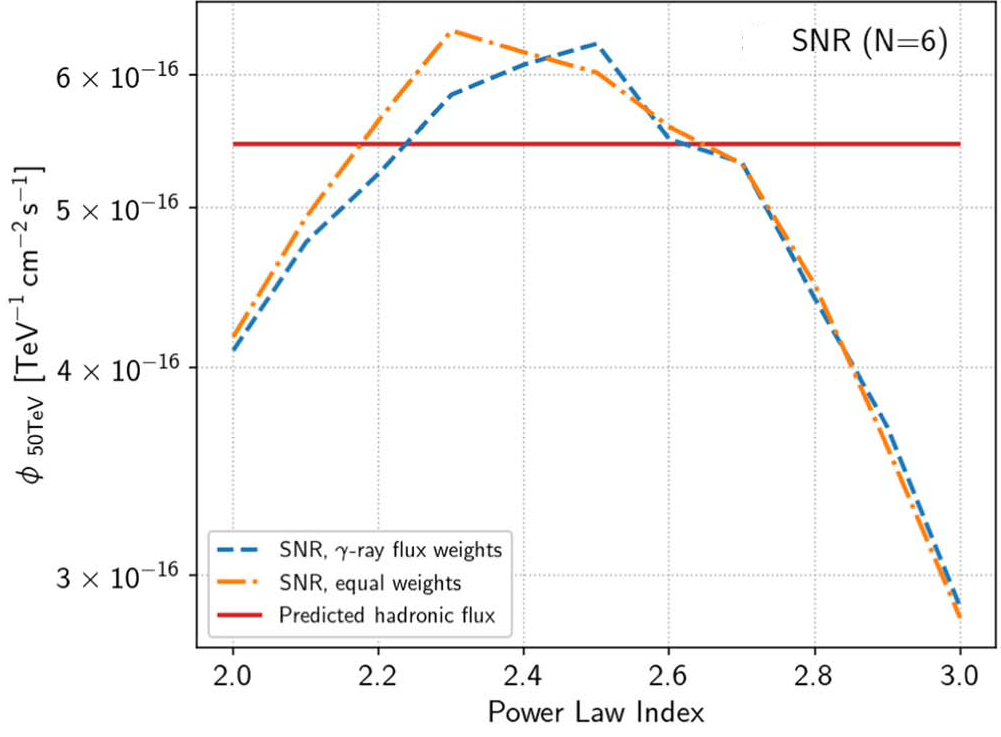}
    \includegraphics[width=0.48\textwidth]{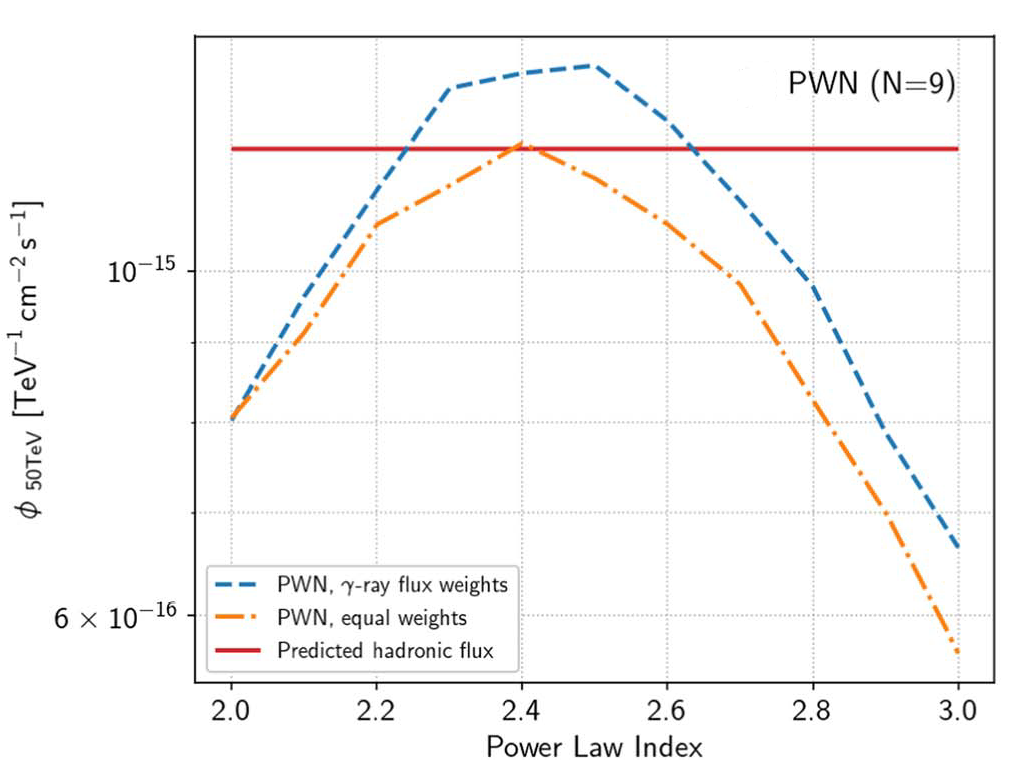}
    \caption{Figures from \cite{Abbasi23_neutrini}, where is indicated the 90$\%$ flux upper limits on neutrino emission at 50 TeV from LHAASO sources that could be associated with SNRs (left) and with PWNe (right). The orange line represents the equal neutrino flux at Earth (the relative weight of the sources is 1/N), and the blue line represents the neutrino flux proportional to the \gray{} flux (relative weight based on the \gray{} flux of every source). The red line is the hadronic flux provided by the assumption that the whole \gray{} flux is of hadronic origin.}
    \label{Fig:Abbasi_neutrini}
\end{figure}

These results were confirmed also by the search for neutrinos from the 12 LHAASO sources with 11 yr of track-like events (neutrino colliding with matter in or near the detector, resulting in a HE muon that travels a long distance) from the IceCube Collaboration \citep{Abbasi23_neutrini}. In this last work, the authors made also a stacking analysis combing: 1) all LHAASO sources, 2) LHAASO sources with SNRs as potential TeV counterpart, 3) LHAASO sources with PWNe as potential TeV counterpart. Even in this case, according the strong assumption that all the flux measured by LHAASO is hadronic, there are no-significant detection but a $90\%$ UL on the predicted hadronic flux. This is an indication that there is a low chance that all 6 sources with a SNR as candidate actually are SNRs is very low and the same is valid for the 9 possible PWNe (see Fig. \ref{Fig:Abbasi_neutrini}).  

In all the works, however, is stressed the strong chance represented by future instruments, as IceCube-Gen2, to detect neutrino emission by LHAASO PeVatrons, in the case that there were hadronic PeVatrons (see Section \ref{Sec:future}).

%!!!!!!!!!!!!!!!! 
%!!!!!!!!!!!!!!!! 

Several theoretical works were developed after the LHAASO results, analyzing which kinds of sources are able to satisfy the requirements needed to explain the CR density in our Galaxies, in spite of the acceleration mechanism at the origin (DSA is the main one but there is the magnetic reconnection that can contribute in microquasar jets or in the Wind of the PWNe) \citep{Morlino22H, Aharonian22H, Gabici22H, Amato21, Olmi23}. One of the main hypotheses so far is that, since the CR spectrum in the energy range around the “knee” shows several features, maybe we can have different contributions from different kinds of sources at different energies \citep[][and Fig.\ref{Fig:YMSCs_contribution}]{Reville22H}.
A parallel analysis of \gray{} and neutrinos from different kinds of sources can give us the right key to interpret the recent results by LHAASO and by the other VHE-UHE \gray{} instruments \citep{Liu21, Abbasi23_neutrini, Huang22_neutrini_a, Sarmah22a}, in order to discern hadronic from leptonic accelerators. This kind of analysis is important also to compare known and well studied sources with the same types of sources candidate PeVatrons, to understand which are the physical conditions at the origin of the differences.
\\
\\
We can summarize a century of studies in the CRs context in this way:
\begin{itemize}
    \item CRs were discovered in 1912 but, despite the enormous efforts done in recent years, the primary three questions about their origin still need to be clarified: what are their sources, how are they accelerated, and how do they propagate?
    \item Two direct evidences were searched in the \grays{} as a signature for sources of Galactic CRs: the 'pion bump' below 100 MeV, due to the $\pi^{0}$ decay, and \gray{} emission at $E>100$ TeV. 
    \item In the standard paradigm, the main candidate for Galactic CR acceleration are SNRs, mainly for energetic reasons. In spite of the evidence of "pion bump" in some of them, no SNRs were detected at E>100 TeV and the hypothesis that other kinds of sources could contribute to CR acceleration made its way.
    \item First observational evidence of VHE emission without the hint of a cut-off from MSCs, microquasars and PWNe confirmed this last hypothesis and they questioned the emission at $E>100$ TeV as the "smoking gun" of hadronic acceleration.
    \item The detection of 12 Galactic Pevatrons by LHAASO confirmed both that sources other from SNRs can accelerate CRs and that electrons can emit UHE \gray{}. Consequently we need a further help to understand which sources are really the CR accelerators.
    \item Neutrino detection and a better morphology reconstruction are fundamental tools for this aim and future instruments can give us a great help. 
   % \item In the meanwhile, further observations with current instruments and development of theoretical model based on the last results are producing interesting and very useful results.
\end{itemize}

Starting from this context, we think that a summary of the work done on the LHAASO sources after the publication of their 12 PeVatron paper \citep{Cao21} could be very useful to the scientific community, and also to future instruments, to understand what could be the right action plan. 
We are aware that other instruments before LHAASO, e.g. HAWC or MAGIC, obtained results that hinted at what LHAASO saw successively. However, we think LHAASO results were a breakthrough for two main reasons: the extension by an order of magnitude of the energy range and the relatively large number of sources.
Despite the fact the portion of the Galaxy observable by the LHAASO site is smaller than that by the HAWC site (and it doesn’t include the Galactic center), LHAASO tripled the number of known sources emitting above 100 TeV, demonstrating that this feature is quite common on galactic sources. The number of sources for which a hadronic scenario is almost excluded increased similarly. 
Moreover, the detection of photons near 1 PeV challenges an astrophysical explanation of what we are observing (both assuming hadronic or leptonic origin).
Let us also note that LHAASO obtained these results with only 10 months of observations, suggesting these 12 sources are only the top of the iceberg of what will be detected by future observations of both LHAASO and other instruments based on the same technology (i.e. SWGO \citep{Schoorlemmer19_SWGO} at the South or TAIGA \citep{Gress17_TAIGA}).

In Section \ref{Sec:sources}, we briefly summarize what we know on the source types that could contribute to Galactic CR acceleration: SNRs (Sect. \ref{Sec:SNRs}), MSCs (Sect.\ref{Sec:MSCs}) and PWNe (Sect. \ref{Fig:PWNe}). We give a look also to TeV halos (Sect. \ref{Sec:halos}) because, even if they are expressions of already accelerated CR diffusion, more than one LHAASO source could be associated with a halo. For completeness, we include in this section also Microquasars (Sect. \ref{Sec:microq}) because, although these kinds of sources are not associated with any LHAASO PaVatron, they are considered between the possible hadronic PeVatrons. In Section \ref{Sec:12PeV}, instead, we focus on each of the 12 LHAASO UHE detections and on the information about them collected up to now. Finally, we conclude with a look on the expectations from future instruments (Section \ref{Sec:future}) and with final remarks (Section \ref{conclusions}). 

%%%%%%%%%%%%%%%%%%%%%%%%%%%%%%%%%%%%%%%%%%
\section{PeVatron candidate source typologies}
\label{Sec:sources}
%%%%%%%%%%%%%%%%%%%%%%%%%%%%%%%%%%%%%%%%%%

Despite some features revealed in the SNRs confirm that these sources accelerate the low energy component of Galactic CRs, we do not have proofs of their UHE \gray{} emission. Even the youngest SNRs, with a high shock velocity, amplified magnetic field and dense environment, show spectra with a cut-off well before $E=100$ TeV (see Fig. \ref{Fig:YSNRs_spectra}). Moreover, there are other challenges correlated to the role of SNRs in CR acceleration; their observed \gray{} spectra, that are different from the theoretical ones, and some problematic issues correlated with CRs abundances \citep[see][and reference therein]{Cristofari21}, motivated the search of alternative sources.

\begin{figure}[!h]
    \centering
    \includegraphics[width=0.6\textwidth]{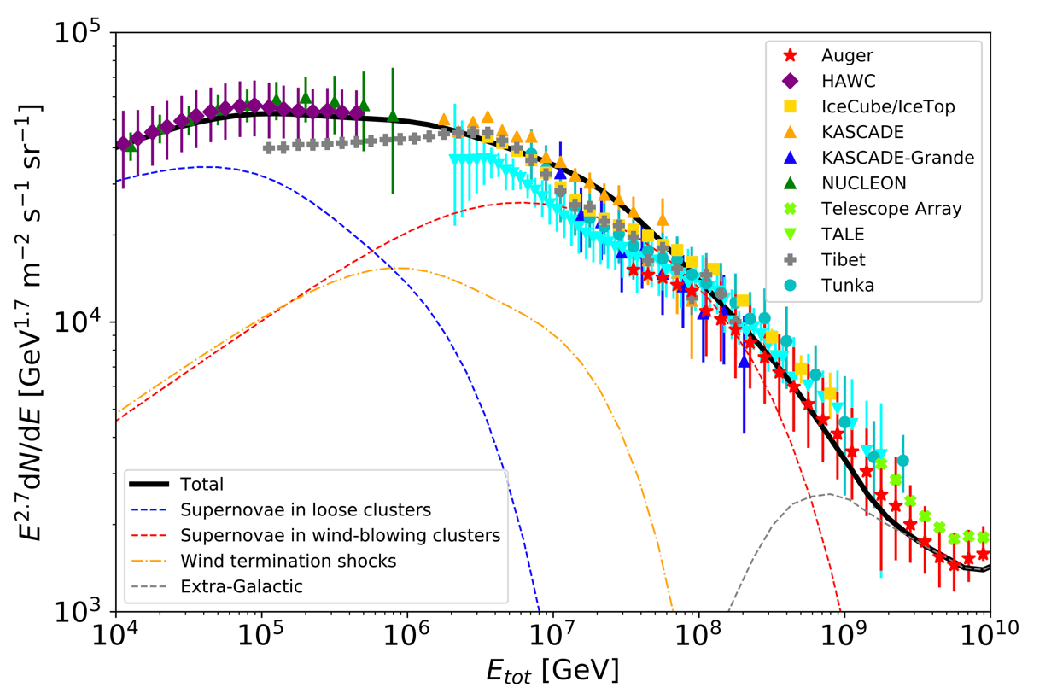}
    \caption{Figure from \cite{Vieu23}. Here, the all-particle CR spectrum is explained with the contribution from different kinds of Galactic sources. In particular, it is evident the dominant contribution from the SNRs in MSCs.}
    \label{Fig:YMSCs_contribution}
\end{figure}

\begin{figure}[!h]
    \centering
    \includegraphics[width=0.5\textwidth]{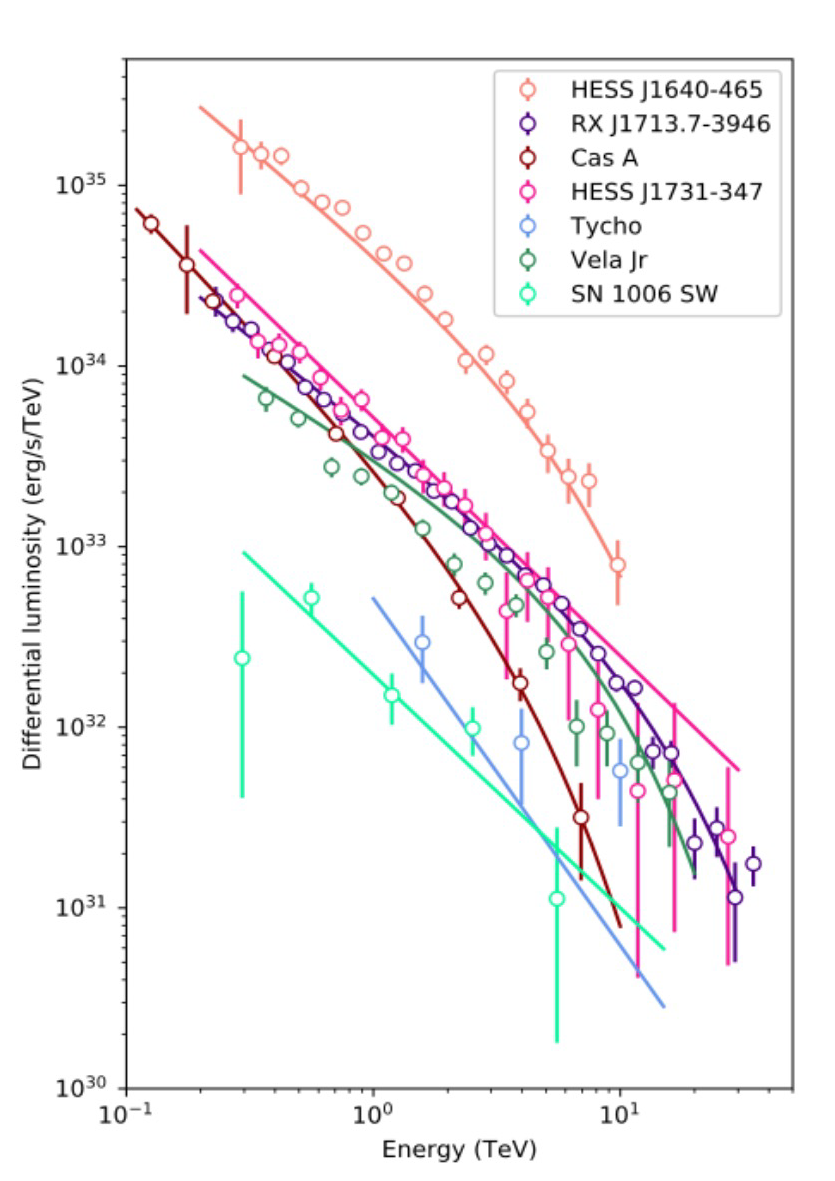}
    \caption{Differential luminosity curves for some of the most commonly studied SNRs at TeV energies. It is evident how, even the youngest SNRs, they cannot reach $E>50$ TeV [Figure from \citep{Aharonian19}].}
    \label{Fig:YSNRs_spectra}
\end{figure}

Furthermore, in the last years the assumption that SNRs may not be the only source of CRs was supported by several detections of VHE/UHE \gray{} emission from different kinds of sources, already before the LHAASO results. This possibility is supported also by the higher complexity of the CR spectrum shown by CALET (see Fig.\ref{Fig:CRs}, right) that seems to be an indication that we need several kinds of sources to explain all the CR spectral features. Indeed, different sources can contribute to the CR spectrum at different energies \citep{Reville22H, Vieu23}: in this hypothesis, for example, the CR 'knee' could be the transition region between the contribution of standard isolated SNRs and the one of the SNRs in MSCs (see Fig. \ref{Fig:YMSCs_contribution}).

%!!!!!!!!!!!!!!!! 
%!!!!!!!!!!!!!!!!

After the large amount of experimental data collected so far and of theoretical models developed in the context of CR acceleration, we have important constraints that a certain kind of source must comply to be considered a hadronic Pevatron.

From a theoretical point of view, the hadronic PeVatron requirements can be summarized in the following way \citep{Morlino22H}:
\begin{itemize}
    \item Energetics - Hadronic PeVatrons must explain the CR power of the Galaxy, estimated to be about $P_{CR}\sim 10^{40}-10^{41}$ erg/s \citep{Gabici19};
    \item Power-law injected spectrum - According to DSA, the accelerated spectrum at the source must follow a power-law that, after correction for transport effects and CR anisotropy, has to be $\propto E^{-2.1\div2.3}$ \citep{Blasi19, Gabici19};
    \item Maximum energy - The acceleration efficiency at the source must be such as to explain PeV particle energies;
    \item Anisotropy - The source distribution must be correlated with the CR anisotropy in their arrival direction predicted by any diffusive model for CRs \citep{Amato14a, Gabici19};
    \item Composition - Galactic CR sources must explain their composition that is slightly different with respect to the Solar CR one (e.g. large overabundance of light elements, overabundance of metals with respect to H and He) \citep{Blasi19, Gabici19}. Moreover, the composition varies over energy (e.g. different elements have different spectral indices), and this also this energy dependence needs to be reproduced. 
\end{itemize}

From an observational point of view, a candidate hadronic PeVatron must have some precise features \citep{Hinton22H}:
\begin{itemize}
    \item Evidence of UHE \gray{} emission ($E>100$ TeV) - If we have CR acceleration, we expected protons accelerated at least up to 1 PeV and consequently, \gray{} emission well above 100 TeV without a cut-off in the spectrum;
    \item Spectral curvature -  The parent particle power-law is modified by the existence of a maximum energy and/or to some break due to losses; 
    \item Correlation with target material - The p-p interaction is most favoured in presence of a dense target in the source surroundings \citep{Blasi13,Amato14a}
    \item Extended emission with likely energy-dependent morphology - CRs diffuse through the media surrounding the source and diffusion (as like as cooling) is energy dependent;
    \item Multi-wavelength counterpart - Secondary non-thermal electrons produced in the p-p interaction emit in radio/X-ray bands.
\end{itemize}

All these observational requirements can be found also in leptonic PeVatrons. Consequently, a cross-check between experimental data and theoretical models is fundamental in order to disentangles PeVatrons and CR sources (hadronic PeVatrons).
\\
\\
In the following, we summarize the classes of sources that could be associated with LHAASO candidate PeVatrons and, consequently, that are also candidate CR accelerators (hadronic PeVatrons): Supernova Remnants, Massive Star Clusters and Pulsar Wind Nebulae \citep[see also][for a brief and good review]{Vink22}. For completeness, we dedicate a paragraph also to TeV halos, the new class of VHE objects that are due to leptonic emission but that could be associated to LHAASO UHE emissions, and to Microquasars, candidate hadronic PeVatrons but without any LHAASO UHE emission associated.

%---------------------------------------------------
\subsection{The standard candidates: Supernova Remnants}
\label{Sec:SNRs}
%---------------------------------------------------
Since the beginning of the CR study, the SNRs have been considered "the sources" of Galactic CRs because they respect most of the theoretical and observational requirements to be hadronic PeVatrons \citep[see][for a recent review]{Cristofari21}. First of all, their power is able to sustain the measured CR flux in our Galaxy: $\dot{E}_{SN}\approx 10^{42}$ erg s$^{-1}$, consequently, $E_{CR}\approx 10\% \dot{E}_{SN}$ \citep{Ginzburg64}. Then, they are ideal laboratories for particle acceleration through DSA mechanism \citep{Fermi49, Berezhko99, Bell15}; first order Fermi Acceleration mechanism works very well in their environment, producing the required pure power-law spectrum. 
Several experimental results have supported this hypothesis with indirect proofs. First of all, several detections of their extended emission at HE and VHE \gray{} and their very frequent association with high-density MCs. Then, the presence of non-thermal and fast variable X-ray emission coincident with TeV \grays{} indicate the presence of magnetic field amplification, fundamental condition to reach VHE energies \citep{Koyama95,Aharonian04, Uchiyama07, Vink12}. The modified Balmer lines detection is an evidence of the CR back-reaction to the DSA \citep{Morlino12} and finally their spatial distribution is compatible with CR distribution and enough SNRs exist to explain the anisotropy \citep{Gabici19}. 
Furthermore, a first real breakthrough came in 2011 when AGILE satellite showed for the first time a direct proof of CR presence in a shock of a SNR, the SNR W44 \citep{Giuliani11, Cardillo14}, then confirmed by Fermi-LAT in the same source W44 and in others similar middle-aged SNRs as IC443, W51c and W49b \citep{Ackermann13,Jogler16_W51c,Abdalla18_W49b}. All these sources were detected at E<100 MeV with a spectrum clearly correlated to the "pion bump" presence.

However, these detections confirmed that SNRs can accelerate CRs of low energies ($E<100$ GeV) but a \gray{} clear detection at UHE is still missing. 
Theoretical works have shown that this could be due to intrinsic properties of SNRs for which they accelerate PeV particles only in the first 100 years of their evolution \citep{Lagage83_EM, Bell13, Cardillo15} and only very rare (1 every $10^{4}$ yrs) high powerful ($E_{SN}\sim 5-10\times10^{51}$ erg) explosions can accelerate CRs to PeV energies at the beginning of their Sedov phase \citep{Cristofari20}. The only chance to detect UHE \grays{} from SNRs is if accelerated CRs, after their escape from the sources, are trapped in near GMCs thanks to a suppression of the diffusion \citep{Gabici07_MCs, Gabici19, DeOnaWilhelmi20,Sushch23}. The LHAASO results show that about six of its 12 PeVatrons could be associated to (middle-aged) SNRs. Consequently, if a deeper study of these UHE emission regions should confirm their SNR nature, we will be able to know if the emission is actually from near GMCs or, in case the UHE emission comes from the SNR itself, if we need to improve our knowledge of the SNRs microphysics \citep{Aharonian22H, Cristofari20}.

Anyway, there are also other difficulties related to the link between SNRs and CR acceleration. Predicted spectrum does not match the observed ones, requiring modifications on the original DSA theory \citep[see][for a recent review]{Blasi19, Cristofari21}. Moreover, if SNRs are the only CR sources, there are several composition issues that are not easily to explain: the most popular example is the $^{22}Ne/^{20}Ne$ abundance, that in Galactic CRs is about 5 times larger than the one in Solar CRs and it cannot explain in a SNR context. In front of these many challenges, the search for other Galactic CR contributors is well motivated.

%!!!!!!!!!!!!!!!!   
%!!!!!!!!!!!!!!!! 

%Vink 2022, Vink HONEST, Safi-Harb HONEST, Giuffrida HONEST, Morlino HONEST, Aharonian HONEST

%---------------------------------------------------
\subsection{Superbubbles and Young Massive Clusters}
\label{Sec:MSCs}
%---------------------------------------------------
The first interest on the role of Superbubbles (SBs, shell structure originating from multiple stars in OB associations) around Young MSCs (YMSCs, $M\gtrsim10^{4}M_{\odot}$) in the CR context was in the early 80's \citep{Cesarsky83}. Indeed, they have several possible acceleration sites, since they host hundreds of massive stars with their fast winds \citep{Montmerle79}, have a turbulent interior \citep{Klepach00} and a lot of potential SNRs expanding inside them \cite{Parizot04}. The entire bubble can reach dimensions greater than 100 pc, creating expectations of extended sources, then confirmed by several VHE detection of regions surrounding young MSCs \citep{HESS18_GPS, Aharonian19}, e.g.  Cygnus Cocoon \citep{Aharonian02_J2032,Ackermann11_Cocoon, VERITAS18_Cygn, Abeysekara21_Cocoon, Amenomori21_Cygnus}, Westerlund 2  \citep{HESS11_West2}, 30 Dor in the Large Magellanic Cloud \citep{HESS15}, the Central Molecular Zone \citep{HESS16_CMZ, HESS18_GC} and Westerlund 1 \citep{Aharonian22_Westerlund1} \citep[a complete list in][]{Morlino22H}. Moreover, the collective effects of shocks from the SN explosions and the Wind Termination shocks in PWNe can explain the mismatch of models to the observed $^{22}Ne/^{20}Ne$ ratio \citep{Gupta20, Gabici22H}, solving one of the main issue linked to the SNR role in CR acceleration \citep[see][for a review]{Bykov20}. 
 
The power injected by a single stellar wind is between 10-50$\%$ of the power generated by a SNR shock \citep{Morlino22H} with an average energy per massive star of about $10^{48}$ erg \citep{Vink22, Vink22H}. Consequently, isolated massive stars fail to supply the CR energy budget in our Galaxy. However, in the most compact clusters (YMSCs with size of a few pc, called also Wind-blowing clusters by \cite{Vieu23}) can merge powerful winds from youngest stars and Wolf Rayet stars into a collective outflow that accelerate CRs in the whole region \citep{Vink22, Vieu23}. The active mechanism should be the less efficient second order Fermi acceleration \citep{Harer23} but this lasts up to million years, much more time with respect to the efficient phase of the SNRs. 

The long duration of CR injection is confirmed also by the spatial \gray{} distribution detected in Cygnus OB2, Westerlund 1 and Westerlund 2 that implies a CR distribution $\propto 1/r$ \citep{Aharonian19}, indication of a continuous injection.
Moreover, magnetic turbulence with $B\sim 100 \mu G$ could be reached inside the core \citep{Badmaev22}. This allows YMSCs to reach E > PeV with a spectrum in agreement with the observations \citep{Morlino21} and to explain CR spectrum at knee energies and above, even if with an expected and strong dependence on the diffusion coefficient \citep{Morlino21, Gabici22H, Morlino22H, Vieu23}. According to these models, the "knee" of the CR particle spectrum can be explained by the transition between the isolated standard SNRs and the SNRs in compact YMSCs (see Fig. \ref{Fig:YMSCs_contribution}).

Summarizing, according to observational and theoretical criteria, YMSCs and Superbubbles may be one of the main contributor of Galactic CRs. However, theoretical models are still incomplete and higher angular resolution of future IACTs together with multi-wavelength and multi-messengers observations will be fundamental in order to resolve the morphology of these regions.

%---------------------------------------------------
\subsection{Pulsar Wind Nebulae: the Crab and the others}
\label{Sec:PWNe}
%---------------------------------------------------
PWNe should be the most numerous between the different kinds of Galactic sources \citep{Fiori22} and their \gray{} emission is estimated to last a very long time ($\sim 100$ kyrs) \citep{Olmi23}. This is likely one of the reason for which more than 30$\%$ of the candidate PeVatrons detected by LHAASO seem to be associated with PWNe \citep{DeOnaWilhelmi22}. They are the most efficient accelerators in the Galaxy (up to $30\%$ of acceleration efficiency), with no signs of thermal particles, and the only sources in which we have direct proof of PeV electron acceleration thanks to their synchrotron spectrum. The location of particle acceleration should be the Termination Shock of the young PWN, generated by the friction between the pulsar wind and the parent SNR. 

With their UHE detection by LHAASO and other EAS experiments, PWNe respect all the observational requirements to be PeVatrons (and possible hadronic PeVatrons);  extended HE \gray{} emission, multi wave-length counter part from radio to X-rays, non-thermal spectrum with spectral curvature \citep{DeOnaWilhelmi22,Olmi23}. The main mechanism producing this UHE emission is the IC scattering of UHE accelerated electrons with Cosmic Microwave Background Radiation (CMBR). From a theoretical point of view, instead, the context is more complex. The first issue is the energetics. At the LHAASO detected energies, the Klein-Nishina regime is the dominant one and, in order to explain leptonic UHE \gray{}, a very high acceleration rate is required \citep{Amato21}. 
%and this give to us the relation between \gray{} photons and parent electrons, $E_e\simeq2.15(E_{\gamma}/1 PeV)^{0.77}$ PeV, and between \gray{} photons and the mean energies of the synchrotron emission, $\epsilon_{syn}=9.3(E_{\gamma}/1 PeV)^{1.5}(B/100\mu G)$ MeV \citep{LHAASO21_Crab}.
Indeed, in spite of acceleration mechanism active in PWNe, still unknown, we know that there are two constraints to the reachable particle maximum energy. The first strong limit is due to the polar cap potential drop: $E_{max}=q(\dot{E}/c)^{1/2}$, where q us the particle charge \citep[see ][for a quick review]{Amato22H, DeOnaWilhelmi22}. In order to have a direct link between theory and observation \cite{DeOnaWilhelmi22}, we can write that relation as: $E_{\gamma,max}\approx0.9\eta_{e}^{1/3}\eta_{B}^{0.65}\dot{E}_{36}^{0.65}$ PeV where $\eta_B$ is the fraction of PWN energy flux transformed in magnetic energy density and $\eta_e$ the ratio between electric and magnetic field. This expression say to us that only a very energetic pulsar ($\dot{E}\gtrsim10^{36}$ erg/s) can reach PeV energies. The second constraint is due to the condition that the acceleration rate must be higher than the electron radiative losses: $E_{\gamma,max}\approx2.7\eta_{e}^{0.65}\eta_{B}^{-0.33}R_{0.1}^{0.65}\dot{E}_{36}^{-0.33}$ PeV, where $R_{0.1}$ is the Termination Shock radius in units of 0.1 pc \citep{DeOnaWilhelmi22}. This condition is more stringent for young energetic pulsars ($\dot{E}\gtrsim10^{37}$ erg/s), as like as the Crab Nebula. 

In this scenario, the presence of accelerated hadrons cannot be excluded but some basic calculation show that PWN cannot be the main CR contributors. In order to produce an amount of energy comparable with the energy expected from SNRs, they should have an initial spin period too small with respect the one provided by theoretical models \citep{Vink22}. This implies that, even if PWN should contribute to Galactic CRs, they have no chances to explain CR composition or anisotropy.

\begin{figure}[!h]
    \centering
    \includegraphics[width=0.8\textwidth]{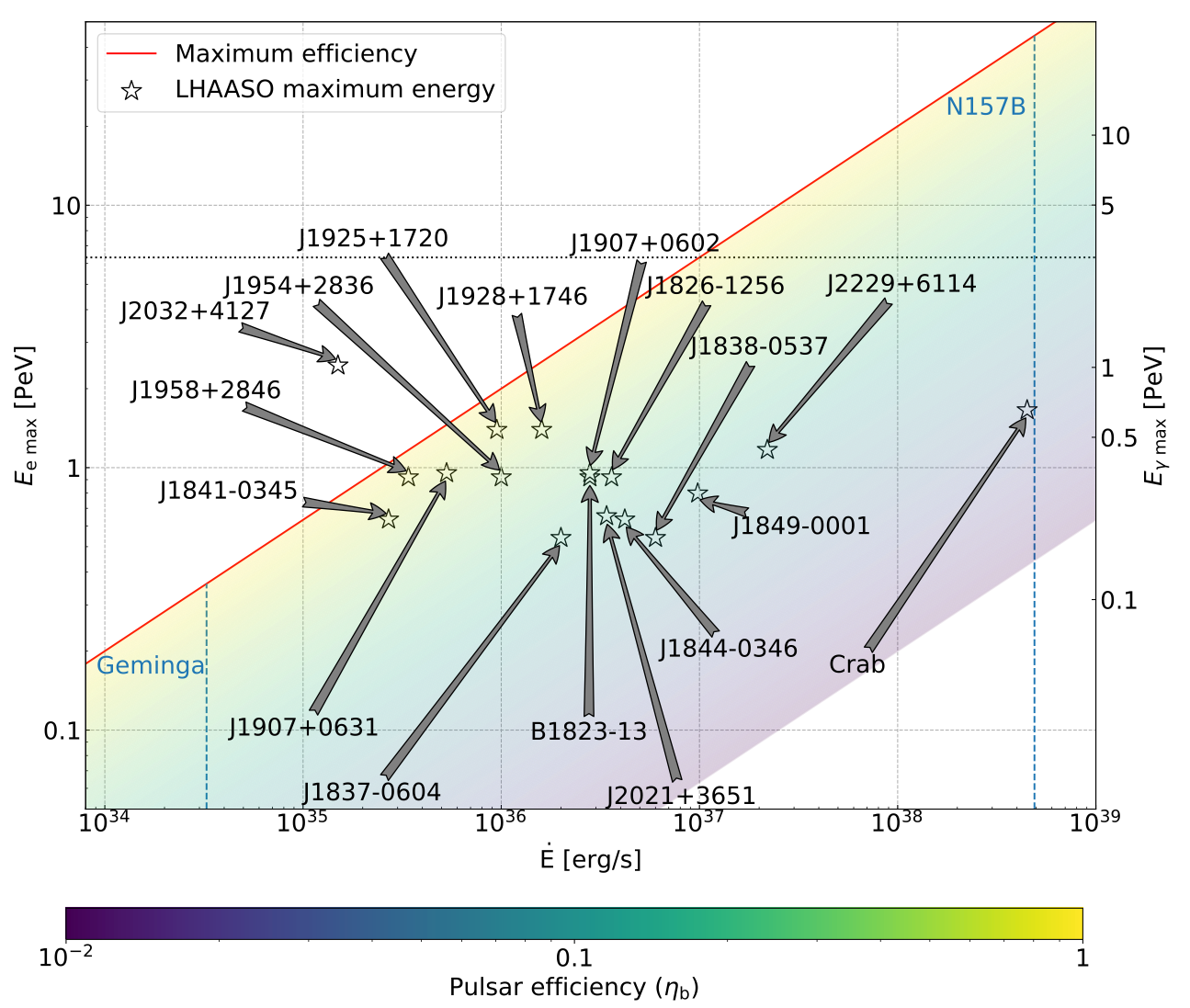}
    \caption{The LHAASO candidate PeVatron with a PWN as potential TeV candidate in the PWN parameter space according to the model in \cite{DeOnaWilhelmi22}. All the sources could be PWN without violating the PWN physics but LHAASO J2032+4127, likely associated with the Cygnus Cocoon [Figure from \citep{DeOnaWilhelmi22}].}
    \label{Fig:PWNe}
\end{figure}

A very good summary of LHAASO PWNe candidate PeVatrons and their characteristics can be found in \citep{DeOnaWilhelmi22} where all the LHAASO sources with a PWN as potential TeV counterpart are analyzed in the PWN parameter space based on energetic constraints. 
According to this work, all these LHAASO \gray{} emissions near a PSR could be produce by a PWN but LHAASO J2032+4127 (see Fig.\ref{Fig:PWNe} and the paragraph dedicate to LHAASO J2032+4102 (\ref{Sec:Cocoon})).

%In particular, the Crab Nebula was detected at \gray{} energy at about 1 PeV, implying a conversion of rotational energy to non-thermal energy with an efficiency of $50 \%$.
%!!!!!!!!!!!!!!!! 
%!!!!!!!!!!!!!!!!

%Vink 2022, Safi-harb HONEST, Giacinti HONEST, Aharonian HONEST, Olmi, Amato HONEST e non solo

%---------------------------------------------------
\subsection{TeV halos}
\label{Sec:halos}
%---------------------------------------------------

The discovery of TeV halos surroundings pulsars can be attributed to the extended TeV emission detected around Geminga and Monogem (B0654+14) pulsars by MILAGRO \citep{Abdo07_MILAGRO_survey} and HAWC \citep{Abeysekara17_2HWC}. The extension of these \gray{} sources is about $2^{\circ}$, indicating that CR propagation near these pulsars is more constrained with respect to what happens in the ISM \citep{Linden17_TeV_halos}. TeV halos can represent, indeed, a local environment for studying CR propagation with small-scale diffusion models \citep{Fang22_TeVHalos}. Their emission is due to a slow diffusion of the electrons in the ISM around the pulsars (several hundred time smaller than the Galactic diffusion coefficient \citep{Aharonian21_J0621}). Although we don't know the nature of the environment that implies a slow down of the diffusion, we know that the \gray{} emission is due to IC scatter of electrons accelerated by the pulsar on the Galactic and cosmological radiation fields \citep{Fang22_TeVHalos}. The discovery of this new class of objects is important also because they could allow to discover a population of "invisible" pulsars that have no beamed radio or HE beamed emission. Moreover, the spectral study of TeV halos is an alternative method to constrain the injection of PWNe electrons because their spectrum depends only by the electron spectral cut-off. 

The large extension ($\sim1$ degree) of many of the LHAASO source and their correlations with pulsars enhance the interest around this new class of sources (see Table \ref{Fig:Cardillo_table}). The only firm TeV halo detected by LHAASO is LHAASO J0621+3755 \citep{Aharonian21_J0621} that is not treated in this review because of its low UHE \gray{} significance ($<5 \sigma$ at 100 TeV). It is one of the currently known TeV halos together with Geminga halo, Monogem halo and HESS J1831-098 \citep{Fang22_TeVHalos}. The main problem correlated with the TeV halo detection is the chance to confuse them with extended SNRs \citep{Linden17_TeV_halos}. In order to verify TeV halo origin of a source, beyond the co-location of the bulk of its emission with a pulsar, this pulsar need to have a spin-down luminosity sufficient to generate a halo and to be enough old to allow electrons to diffuse in the ISM (bow-shock nebula \citep{Olmi23}). Furthermore, the extension of the halo has to be larger than the possible PWN X-ray emission, otherwise the emission could be a \gray{} PWN, and its morphology should be explained by slow-diffusion, otherwise this could indicate a hadronic accelerator \citep{Fang22_TeVHalos}.

The low angular resolution of the EAS experiments doesn't allow to resolve TeV halo morphology and, unfortunately, the size of these object is a challenging for current IACTs because of their smaller Field of View. Geminga and Monogem, indeed, are not detected by these instruments despite their high TeV flux. The future ASTRI Mini-Array, with a FoV of $10^{\circ}$ at E > 10 TeV, and CTA SSTs in the Southern Emisphere, with a FoV $>8^{\circ}$, could finally resolve the morphology of some of these objects and detect a high resolution spectrum. In the ASTRI Mini-Array core Science paper \citep{Vercellone22} a first simulation of the Geminga halo was developed, building a radial profile and showing a high significance spectrum up to 50 TeV with a single observation. Multiple observations will allow the ASTRI Mini-Array to better track the source morphology.

%---------------------------------------------------
\subsection{Gamma-ray binaries and Microquasars}
\label{Sec:microq}
%---------------------------------------------------
A first interest to \gray{} binary systems (i.e. microquasars) like Galactic CR accelerators was born from their similarities to AGN that are likely the main responsible of extra-galactic CR component \citep{Vink22}. We know that microquasars accelerate electrons to VHE \citep{Aharonian05, Khangulyan22H}: their environment is radiation dominated making the Klein-Nishina losses ineffective to prevent acceleration and $\gamma-\gamma$ absorption is transparent for 100 TeV \grays{}. However, they are much less abundant with respect to other kinds of candidate PeVatrons, as SNRs or PWNe. Consequently, even if they could accelerate protons, they cannot explain either CR energy density or anisotropy and composition in our Galaxy. 

This does not mean that some microquasars cannot be hadronic accelerators. So far, several \gray{} binaries system were detected at VHE with no hint of cut-off above 10 TeV: PSR B1259-63/LS2883 \citep{HESS20}, LS 5039 \citep{HESS15,Huang22_LS5039}, V4641 Sagittarii \citep{Tibolla23_HAWC} and SS 433 \citep{HAWC18_SS433, Fang20_SS433, Olivera-Nieto22_Gamma, SafiHarb22_SS433}. In particular, SS443 is detected at E>10 TeV  with a jet that exceed 10 times the required injection power, $L_{kin}=10^{39}$ erg/s \citep{Aharonian22H, Khangulyan22H}. The microquasar V4641, a black hole binary with ultrarelativistic jets, introduces a new class of \gray{} source of strong interest for future instruments. In general, detected emission seems to be dependent by the orbital phase, confirming that should be produced on binary system scale and that there is some acceleration mechanism active. One of the open issue concern if it is the same mechanism seen in the Crab Nebula or if we need to add some typical mechanism active in binary system as like as is suggested by recent X-ray analysis of the SS 433 \citep{SafiHarb22_SS433}.

No microquasars are between the 12 PeVatrons candidate of LHAASO but this class of sources represents an important target for LHAASO and future EAS and IACTs instruments.

%%%%%%%%%%%%%%%%%%%%%%%%%%%%%%%%%%%%%%%%%%%
\section{Overview of the 12 LHAASO sources}
\label{Sec:12PeV}
%%%%%%%%%%%%%%%%%%%%%%%%%%%%%%%%%%%%%%%%%%%%%

%!!!!!!!!!!!!!!!! 
\begin{figure}[!h]
    \centering
    \includegraphics[width=1.06\textwidth]{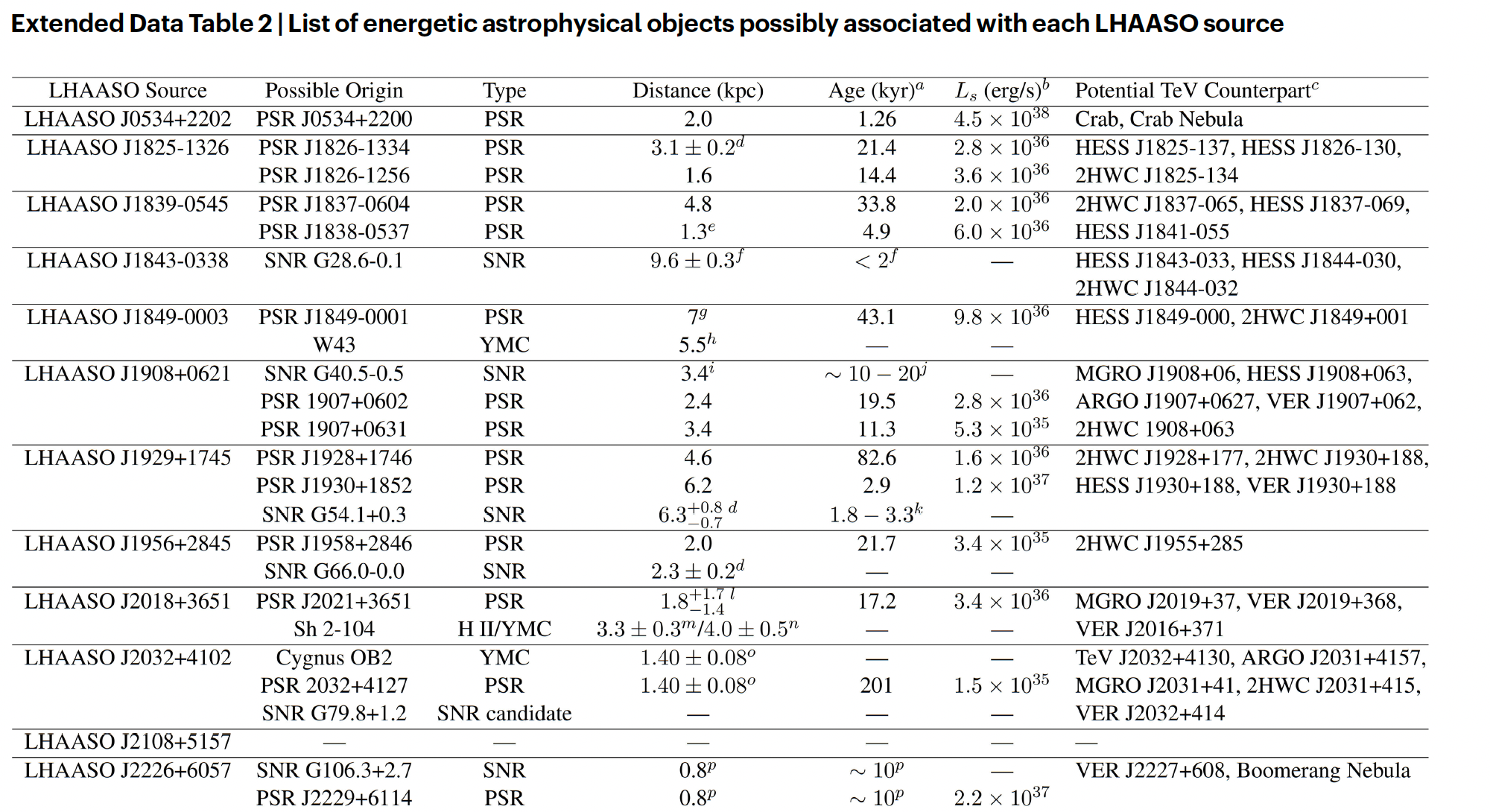}
    \caption{[Figure from \cite{Cao21}] The very popular table with the list of energetic astrophysical objects possibly associated with each LHAASO source. Other information as coordinates, maximum energy and significance can be found always in \cite{Cao21} but also in the Table \ref{Fig:Cardillo_table}.}
    \label{Fig:LHAASO_table}
\end{figure}
%!!!!!!!!!!!!!!!!

The Large High altitude Air Shower Observatory (LHAASO), located in Sichuan (province of China) at 4410 m of altitude, is a complex of EAS detectors with the aim to study both CRs and \grays{} in the sub-PeV up to 1000 PeV energy range \citep{Cao10}. It is composed by three arrays: the Kilometer Square Array (KM2A, 10 TeV - 100 PeV, UHE \gray{} detector with energy resolution $\leq20\%$, angular resolution of $0.25^{\circ}$ and sensitivity (at 50 TeV) of $10^{-14}$ erg cm$^{-2}$ s$^{-1}$), the Water Cherenkov Detector Array (WCDA, down to 0.1 TeV, angular resolution of $0.2^{\circ}$ and sensitivity ($E>2$ TeV) of $10^{-12}$ erg cm$^{-2}$ s$^{-1}$), and the Wield Field-of-view Cherenkov Telescope Array (WFCTA, 0.1-1000 PeV) \citep{LHAASO21_Crab}.  

The 12 Galactic PeVatrons were detected by the half-completed LHAASO-KM2A, with and angular resolution of $0.4^{\circ}$ and an energy resolution of $28\%$ at 30 TeV \citep{Cao21} (see Table in Fig. \ref{Fig:LHAASO_table}). The indicated $E_{max}$ in the Table represents the maximum energy detected for each source (for the Crab Nebula there is an updated value, see the dedicated Section LHAASO J0534+2202 (\ref{Sec:Crab})). Between them, UHE emission in the regions of LHAASO J1825-1326, LHAASO J1908+0621 and LHAASO J2018+3651 was already detected by HAWC \citep[][eHWC J1825-134, eHWC J1907+063, eHWC J2019+368 ]{Abeysekara2020_56TeV}, as like as LHAASO J0534+2202 (Crab Nebula) was detected at $E>150$ TeV by HAWC and Tibet AS$\gamma$ \citep{Abeysekara19_Crab,Amenomori19_Tibet_Crab}.
LHAASO J0534+2202 is the only candidate PeVatron detected by LHAASO with a firm association. The low angular resolution of the LHAASO instruments, indeed, implies that for every  candidate there are at least two candidate TeV counterparts.

In addition to the popular 12 PeVatrons published in \cite{Cao21}, the LHAASO collaboration published two papers on two other sources detected at VHE: LHAASO J0341+5258 \citep{Cao21_J0341} and LHAASO J0621+3755 \citep{Aharonian21_J0621}.
LHAASO J0341+5258 is a unidentified source associated with GMCs \citep{TevCat08} but without any detected PWN or SNR in the surroundings.
LHAASO J0621+3755 is one of the four TeV halos of our Galaxy (see Sec. \ref{Sec:halos}) detected and identified up to now \citep{Fang22_TeVHalos, Martin22_halos}. It is a new source in the TeV domain \citep{TevCat08} and it is spatially coincident with the Fermi-LAT PSR J0622+3749 that has a spin-down sufficient to explain the halo luminosity. Moreover, its spectrum can be explained with super-diffusion models \citep{Fang21_J0621}: all the criteria to be a TeV halo are respected \citep{Fang22_TeVHalos}.

These two sources are not in the main paper \cite{Cao21} and in the PeVatron Table in Fig. \ref{Fig:LHAASO_table} because they have a significance at 100 TeV $<5\sigma$ and for this reason we do not take into account them in this work, where we summarize what have been done from theoretical and experimental point of view on every LHAASO PeVatron source in \cite{Cao21}. Our aim is to clarify what we know and what we have learned about every single LHAASO source and its origin, both from \gray{} and neutrino point of views. A theoretical interpretation and/or analysis of existent models is beyond the aim of this work.

In the table in Fig.\ref{Fig:Cardillo_table}, we summarize the main conclusions about the origin of \gray{} emission from the LHAASO sources together with their coordinates, detected maximum energies and detection significance.

%................
\paragraph{LHAASO J0534+2202 (Crab nebula)}
\label{Sec:Crab}
%................

LHAASO J0534+2202 is the only LHAASO candidate PeVatron with a specified counterpart, the Crab Nebula. LHAASO detected its emission with the highest significance, $\sigma=17.8$. Its maximum energy was about $E_{M} \approx900$ TeV in its first detection by KM2A and WFCTA \citep{Cao21} but, in a second shower detected by KM2A one year later, this reached $E_{M}=1.12\pm0.09$ PeV \cite{LHAASO21_Crab}.

The Crab Nebula is one of the most studied sources of our Galaxy in every electromagnetic band, from radio to \gray{} \citep[see][for a recent and complete review]{Amato21}. In the HE band, its flaring nature at MeV energies \citep{Tavani11,Abdo11_Crab, Yeung20_Crab} was a breakthrough for all the theoretical models but even its steady VHE \gray{} emission is very interesting. Indeed, its synchrotron emission was one of the first evidence of the presence of PeV electrons in its environment \citep{Atoyan96} and its Pevatron nature is confirmed by HAWC \citep{Abeysekara19_Crab} and Tibet AS$\gamma$ \citep{Amenomori19_Tibet_Crab} results that detected it at $E>150$ TeV: only the sensitivity of water and air shower detectors can reach these extreme energies. The very high sensitivity at UHE of the LHAASO KM2A ($\sim10^{-14}$ erg cm$^2$s$^{-1}$ at $E>50$ TeV in 1 yr) showed, for the first time, the \gray{} component of its spectrum extended up to 1.1 PeV, transforming it in a super-PeVatron \citep{LHAASO21_Crab}. Following what we said in Sect.\ref{Sec:PWNe}, the Crab Nebula is one of the youngest and powerful PWN and, consequently, the maximum energy of its \gray{} emission is radiation limited. This means that its detection at 1.1 PeV implies an acceleration rate $\eta_e\approx0.16$, three order of magnitude greater than the one in SNRs \citep{Cao21, LHAASO21_Crab}.

After the LHAASO detection, the main issue linked to the Super-Pevatron nature of the Crab Nebula is related to the acceleration mechanism but also to the chance of a contribution from a hadronic component. Even before this discovery, the chance of a contribution from hadrons was investigated \citep[see][and reference therein]{Nie22_Crab} but now, around 1 PeV the LHAASO spectrum shows an hardening of the spectrum that is really challenging to explain with a leptonic component \citep{LHAASO21_Crab}. The LHAASO error bars are very large and, consequently, this hardening needs to be confirmed but, if it will be, it supports the idea of a UHE hadronic component: a component that cannot explain the overall \gray{} luminosity but that can have a no negligible contribution at the highest energies, with the condition of proton confinement in the nebula \citep{LHAASO21_Crab}. Very recent models estimate that this hardening is compatible with a quasi-monochromatic distribution of protons around 10 PeV \citep{Amato21, Olmi22H} to be added to the confirmed leptonic one. Lorentz factor and the spin-down luminosity carried by protons are two of the most important parameters that characterize this component \citep{Vercellone22}.

\begin{figure}
    \centering
    \includegraphics[width=0.7\textwidth]{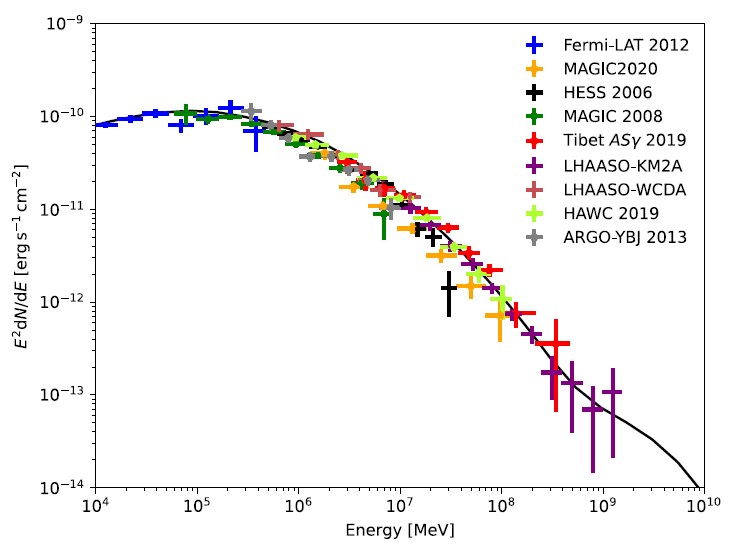}
    \caption{LHAASO J0534+2202. \gray{} spectrum of the Crab nebula with all data points collected up to now and fitted with one of the models that takes into account the contribution of a hadronic component at the highest energies [Figure from \citep{Nie22_Crab}].}
    \label{Fig:Crab_model}
\end{figure}

Several theoretical models to constrain this hypothetical hadronic component are beeing developed \citep{Nie22_Crab, Amato22H} (see Fig. \ref{Fig:Crab_model}) but future instruments as ASTRI Mini-Array and CTA will be fundamental to confirm a lepto-hadronic model for the \gray{} emission from the Crab Nebula. In the ASTRI Mini-Array core science paper \citep{Vercellone22}, a simulation of a Crab spectrum up to 300 TeV was produced, showing how the ASTRI Mini-Array it could be decisive in constraining the likely hadronic component at the highest energies. The Crab Nebula is even the only source for which the neutrino analysis carried on with different methods \citep{Huang22_neutrini_a, Huang22_neutrini_b, Abbasi23_neutrini} can put constraints confirming that the hadronic component is not the dominant one to explain its \gray{} emission. In \cite{Sarmah23_neutrini} was showed that the expected neutrino flux from the Crab is lower than the IceCube sensitivity also in case of pure hadronic emission.

Anyway, the Crab Nebula is a very unique source either for the fact that it is the only PWN where we know that its internal synchrotron radiation is the target of IC and either because, since its young age, its energy losses is dominated by radiation reaction and not by the potential drop as most of the others PWNe \citep{Amato21,DeOnaWilhelmi22}. This is likely the reason for which flaring emission was seen only in the nebula of the Crab.

%!!!!!!!!!!!!!!!! 
%!!!!!!!!!!!!!!!!

%................
\paragraph{LHAASO J1825-1326}
\label{Sec:J1825}
%................
%Safi HONEST, Aharonian HONEST, Burgess22, Deonawilhelmi22

\begin{figure}[!h]
    \centering
    \includegraphics[width=0.45\textwidth]{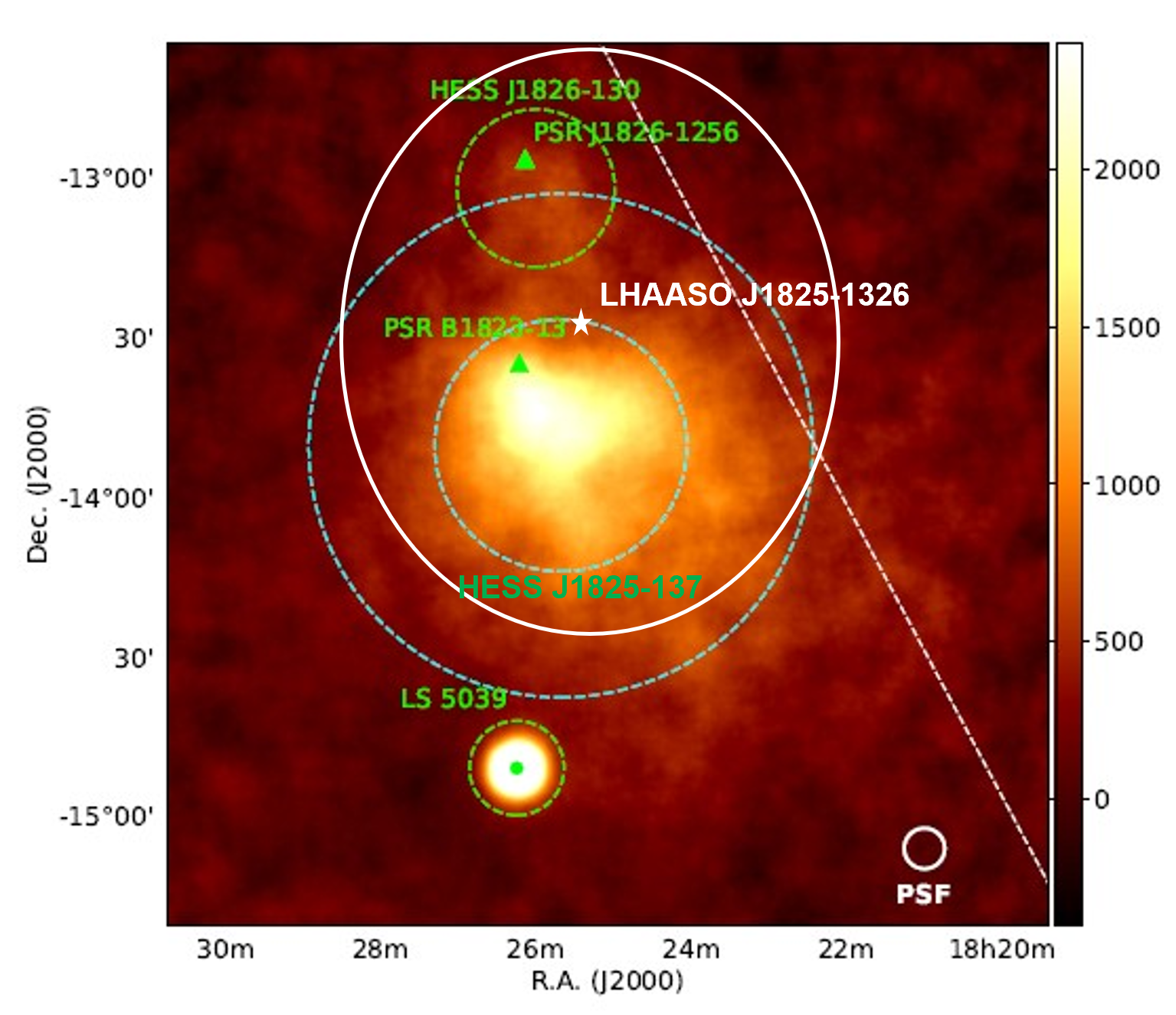}
    \includegraphics[width=0.43\textwidth]{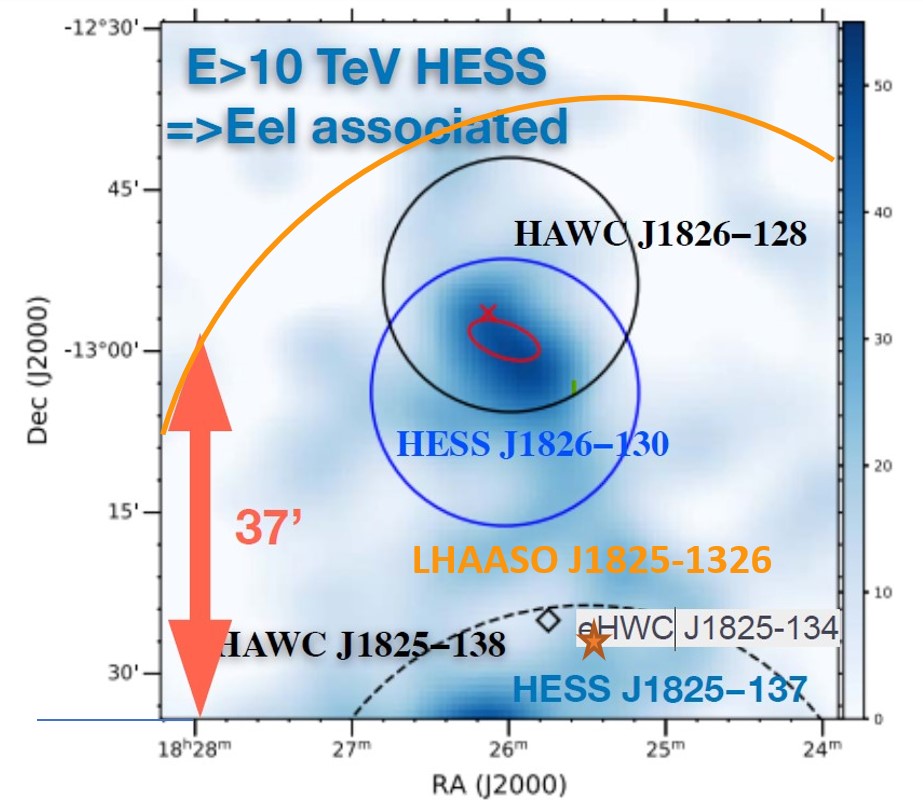}
    \includegraphics[width=0.65\textwidth]{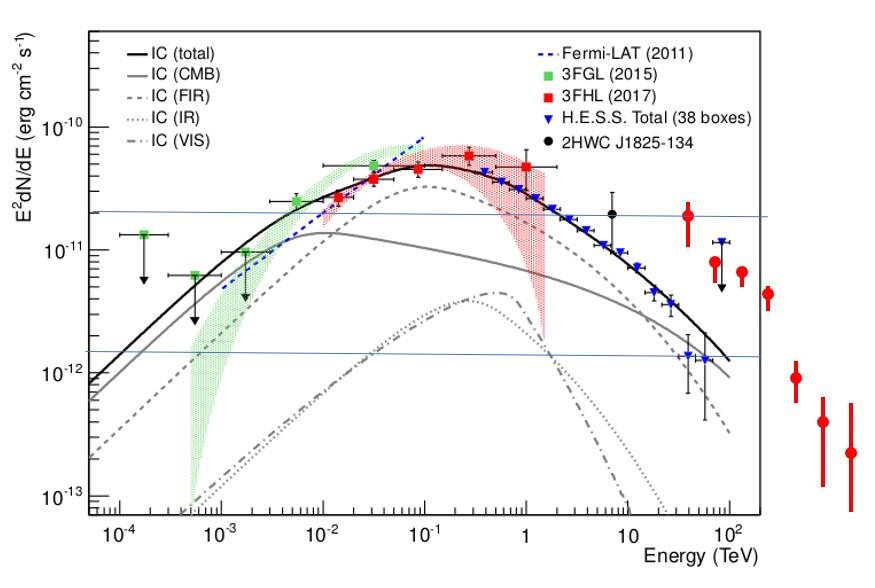}
    \caption{LHAASO J1825-1326. Top Left: HESS excess counts map from \citep{HESS19_J1825} re-adapted with the addition of the LHAASO centroid and a rough estimation of the extension of the LHAASO detection from the image in \cite{Cao21} (white circle). Top right: HESS 1-10 TeV excess map from \cite{Burgess22_J1826} re-adapted by \citep{SafiHarb22_SS433,Cao21} and with the addition of the LHAASO centroid and its extension from \cite{Cao21} (orange circle). The red ellipse indicates the Eel PWN. In this way is evident the reason for which the association of the LHAASO emission is so challenging. Bottom: the spectrum of HESS J1825-137 from \cite{HESS19_J1825} with the LHAASO data points from \cite{Cao21} overlapped.}
    \label{Fig:J1825}
\end{figure}

The candidate PeVatron LHAASO J1825-1326 is one of the sources (the other are the Crab, LHAASO J1908+0621 and LHAASO J2226+6057) detected with the highest significance at E>100 TeV ($\sigma=16.4$) and for which there is a UHE SED without any hint of a cut-off (see Fig. \ref{Fig:J1825}, bottom panel, and paper \cite{Cao21}). Its maximum energy is $E_{M}\approx420$ TeV  and the gradual steepening at the highest energies is maybe due to photon-photon absorption \citep{Cao21}. Its potential TeV counterparts are HESS J1825-137/2HWC J1825-134 and HESS J1826-130, both PWNe in a region smaller than the LHAASO angular resolution (see Figure \ref{Fig:J1825}, top panel).

HESS J1825-137 is the name of the PWN associated to the PSR B1823-13, revealed by the first Galactic Plane Survey of HESS \citep{Aharonian05_J1825} and is one of the brightest and extended sources in the Galaxy at the VHE, with an intrinsic diameter of about 100 pc \citep{HESS19_J1825, Albert21_J1825}. This was detected also by HAWC as a unique extended region, eHWC J1825-134 \citep{Abeysekara17_2HWC}, then resolved in three different \gray{} emitting components  (HAWC J1825-138, HAWC J1826-128 and HAWC J1825-134). In particular, HAWC J1825-134, is at only 0.03$^{\circ}$ by the LHAASO source (see Fig. \ref{Fig:J1825} top right) and its spectrum extends beyond 100 TeV without a cut-off whereas HAWC J1825-138 seems to confirm the spectral behavior of HESS J1825-137 \citep{Albert21_J1825} with a cut-off below 100 TeV. We cannot exclude that also HESS J1825-137 is actually the composition of two different TeV sources.
This HESS \gray{} source shows a strong morphology dependence at GeV and TeV energies \citep{Aharonian06_J1825, Principe19_J1825} and the asymmetry of the nebula in the X-ray band \citep{Gaensler06_J1825} could be due or to the reverberation phase (interaction of the Termination Shock with the parent SNR reverse shock) and/or to the presence of a MC in the surroundings \citep{Castelletti12_J1825}. The extraordinary extension of this source makes it a possible candidate TeV halo \citep{Sudoh19_TeV_halos}, even because it is not possible to explain it within the standard dynamics of the PWN \citep{Khangulyan18_J1825}.

Another between the three HAWC resolved sources, HAWC J1826-128, is detected up to 100 TeV and it is spatially coincident with HESS J1826-130 \citep{HESS18_GPS, HESS20_J1826}, the other potential TeV counterpart of the LHAASO PeVatron candidate, and both the sources seem to be related with the  “Eel” PWN11 (PWN G18.5-0.4) \citep{Burgess22_J1826}. HAWC J1826-128 was detected in X-ray and \gray{} but not in radio band, a strange peculiarity compared to other similar PWNe. In the same field there is also the very poorly known SNR G18.45-0.42 \citep{Karpova19_J1826} that can not be ruled out as the HESS J1826-130 counterpart. According to developed models, a leptonic nature of this emission is favored \citep{Burgess22_J1826}. 

The LHAASO angular resolution cannot resolve the region at the origin of the detected emission, consequently we cannot exclude that the Pevatron emission is due to HESS J1826-130 instead that to HESS J1825-137. Actually, the spectrum of LHAASO J1825-1326 it is one order of magnitude above the spectrum of HESS J1825-137 \citep{HESS19_J1825} and the associated HAWC J1825-138 \citep{Albert21_J1825} (see Fig.\ref{Fig:J1825}, bottom). This could be due to the fact that all three resolved HAWC sources (HAWC J1825-138, HAWC J1826-128 and HAWC J1825-134) can contribute to the LHAASO TeV region (see their VHE spectra shown in \cite{Albert21_J1825}). This implies that a more deep analysis of this TeV \gray{} bright LHAASO region is necessary. It is clear that the better angular resolution of ASTRI Mini-Array and CTA will be fundamental in order to resolve this interesting region.
%but, according to the current information, an association between LHAASO J1825-1326 and HAWC J1826-128/HESS J1826-130 is favoured.

In the IceCube neutrino analysis carried on by \cite{Huang22_neutrini_a} (see Sect. \ref{Sec:A.L.}), LHAASO J1825-1326 is constrained to be leptonic dominant up to 200 TeV. The bayesian approach used instead in \cite{Huang22_neutrini_b}, cannot constrain this source, stressing how these neutrino-based estimation are strongly biased by statistical methods. 

%There is also a study on the possibility to associate this LHAASO emission not to the a nebula but to one of the two PSRs in the field, PSR J1826 −1334, and PSR J1826 −1256 \citep{Chang22,Chang23} 

%!!!!!!!!!!!!!!!! 
%!!!!!!!!!!!!!!!! 

%................
\paragraph{LHAASO J1839-0545}
\label{Sec:J1839}
%................
%Huang22, HAWC J1837-065: HAWC arXiv2019, arXiv:1909.08609, HESS J1837-069: https://www.mpi-hd.mpg.de/hfm/HESS/pages/home/som/2008/05/,  HESS J1841-055: https://arxiv.org/abs/2007.09321, MAGIC22 (Saha et al), Katsuta17, Banik21

LHAASO J1839-0545 was detected by LHAASO with an $E_{M}\approx 200$ TeV in a very complex region. The two challenging potential TeV counterparts are HESS J1841-055 and 2HWC J1837-065/HESS J1837-069/MAGIC J1837-073 \citep{Cao21, TevCat08}. 

The first one, HESS J1841-055, was revealed by HESS in 2007 as an extended TeV source ($\sim0.4^{\circ}$)\citep{Aharonian08_HESS_unidentified}, and then confirmed by HAWC \citep{Abeysekara17_2HWC} and ARGO \citep{Bartoli08_ARGO_J1841} but it hasn't a counterpart and up to now remains an unidentified source. In 2022, the MAGIC collaboration confirmed a TeV detection in coincidence of this HESS detection, finding some hot-spots in the extended emission that could indicate the presence of different sources in the same region that cannot be resolved so far \citep{MAGIC20_J1841, MAGIC22_J1841}. Their analysis consider plausible both PWN and SNR (the middle-aged G26.6-01) origin of the TeV emission, since the presence of a dense MC in the surroundings. The produced spectrum through Fermi-LAT and MAGIC data analysis, however, extends only up to E<10 TeV with an evident cut-off at larger energies. Consequently, this hint could exclude the association of LHAASO J1839-0545 with this complex source. 

On the contrary, HESS J1837-069 is more interesting for the interpretation of the LHAASO emission. This was detected firstly by HESS during its Galactic Plane Survey \citep{TevCat08} and then by HAWC \citep{Abeysekara17_2HWC} and MAGIC \citep{MAGIC19_J1837} and is postulated to be associated with the PWN of the PSR J1838-0655 \cite{TevCat08}. A very detailed study of that region by MAGIC collaboration at GeV-TeV energies \citep{Banik21_J1837} found that the TeV emission in coincidence of the HESS J1837-069 can be interpreted as hadronic emission due to runaway CRs from that source with a MC in the location of MAGIC J1837-073. In this model, a maximum energy up to 70 TeV is considered reachable; LHAASO detected this source up to 200 TeV, giving maybe an important boost forward that interpretation. 

Anyway, there is a further interesting hypothesis. In 2017, a deep analysis of Fermi-LAT data detected from the G25 region (the region that include also the two HESS sources) seems to indicate that there is an extended OB association, G25.18+0.26, with very similar characteristics and behavior of the Cygnus Cocoon region \citep{Katsuta17_J1837} (see Paragraph dedicated to LHAASO J2032+4102 (\ref{Sec:Cocoon})). The spectrum from this region doesn't show any bending and could be a candidate for the TeV emission detected in this region. The very low LHAASO angular resolution cannot allow to exclude also this possible association that could be another confirmation of the strong role of the MSCs in the CR acceleration.

As shown in \citep{Banik21_J1837}, the future IceCube-Gen2 will have the right sensitivity in order to detect possible neutrino emission from this source, confirming or excluding the hadronic origin of this TeV emission.

%Nessuna figura perché non ci sono SED che considerano anche LHAASO e non sappiamo bene cosa associare a LHAASO stesso. La SED in MAGIC22 è solo per la 1841 e non considera giustamente il punto LHAASO

%................
\paragraph{LHAASO J1843-0338}
\label{Sec:J1843}
%................
%Amenomori22,  Castelletti17, Petriella19
%NON ESISTE IL PDF --> Son22:https://ui.adsabs.harvard.edu/abs/2022APS..APRG13007S/abstract
%McCall18 (HESS J1844 come gamma-ray binaries ma non c'è PDF),

\begin{figure}[!h]
    \centering
    \includegraphics[width=0.48\textwidth]{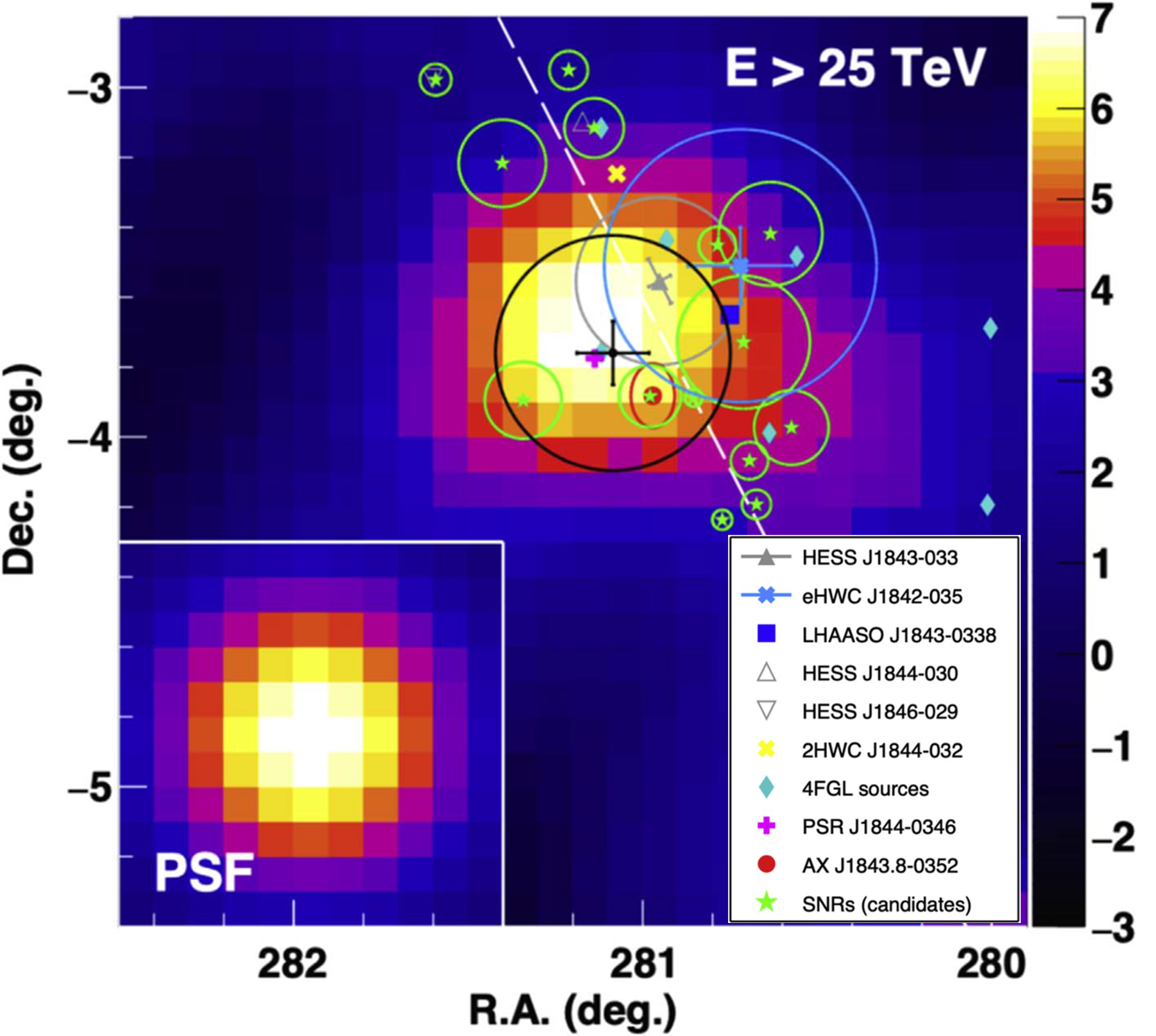}
    \includegraphics[width=0.51\textwidth]{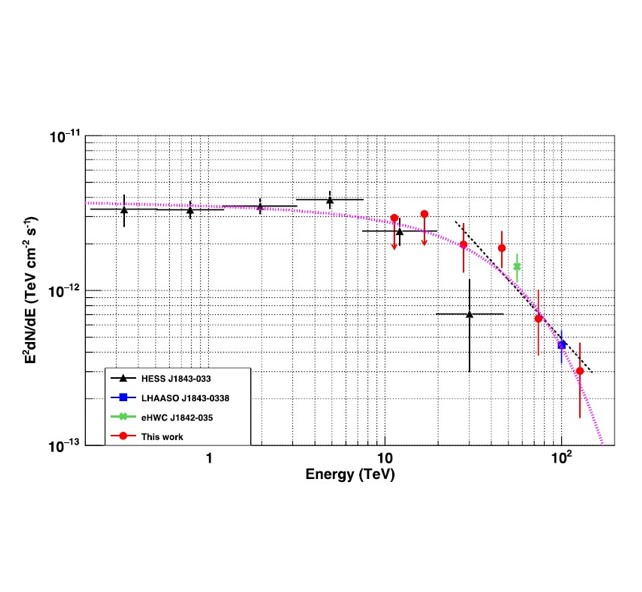}
    \caption{LHAASO J1843-0338. Left: Tibet AS$\gamma$ significance map of TASG J1844-038 at E>25 TeV [Figure from \cite{Amenomori22_J1843}]. Right: the SED of TASG J1844-038 assuming that also the LHAASO data point is correlated with this source [Figure from \cite{Amenomori22_J1843}].}
    \label{Fig:J1843}
\end{figure}

LHAASO J1843-0338, detected with an $E_{M}\approx260$ TeV, is in a very complex region (see Fig. \ref{Fig:J1843}, left) and the association problem is also for the two HESS VHE detections, HESS J1844-030 and HESS J1843-033, candidates as TeV counterparts.

The first source, HESS J1844-030, is a faint point-like VHE source, spatially coincident with a number of distinct objects. At the beginning, it was reported as a component of HESS J1843-033 \citep{Hoppe08} and then was disentangled in the HESS Galactic Plane Survey \citep{HESS18_GPS}. In particular, this emission turned out to be spatially coincident with the radio source G29.37+0.1 that has a very complex morphology. Later, a deep study developed in \cite{Castelletti17_J1844} concluded that this radio source is the superposition of a radio galaxy, a composite SNR with a shell, a pulsar powered component and maybe also a neutron star. At the same time, the authors cannot explain if the nature of TeV \gray{} is produced in the lobe of the radio galaxy or in the SNR/PWN environment. A further analysis, instead, explain all the multi-wavelength emission from the G29.37+0.1 (radio, X-ray and \gray{}) as due to PWN/SNR interaction, excluding the extra-galactic counterpart \citep{Petriella19_J1844}.

The centroid of the LHAASO source is more coincident with the second potential TeV counterpart, eHWC J1842-035/HESS J1843-033. This is an unidentified source discovered by HESS \citep{HESS18_GPS} with a spectrum extended up to 30 TeV. Its complex morphology could be a hint of the presence of overlapping sources merged in the Galactic Plane Survey. In TevCat this is associated with ARGO J1841-0332 \citep{DiGirolamo17_ARGO} and 2HWC J1844-32 \citep{Abeysekara2020_56TeV} and Tibet As$\gamma$ array recently detected a TeV emission above 25 TeV near the position of HESS J1843-033, TASG J1844-038 \citep{Amenomori22_J1843}. The Tibet-AS$\gamma$ emission has a measured spectrum above 100 TeV in perfect agreement with the LHAASO results \citep{Amenomori22_J1843} (see Fig. \ref{Fig:J1843}, right). The authors discussed the origin of this emission, concluding that both the associations with the SNR G28.6-0.1, filled with non-thermal X-ray \citep{Ueno03_J1843}, and PSR J1844-0346, can be considered valid. 

The complexity of the region needs of a VHE/UHE morphology reconstruction in order to disentangled all the \gray{} sources existing there. No neutrino contraints are present so far.

%!!!!!!!!!!!!!!!! 
%!!!!!!!!!!!!!!!! 

%...............
\paragraph{LHAASO J1849-0003}
\label{Sec:J1849}
%................
%Mitchell22: https://pos.sissa.it/395/922/pdf
%http://tevcat.uchicago.edu/?mode=1;id=188
%http://tevcat.uchicago.edu/?mode=1;id=345

LHAASO J1849-0003 was detected by LHAASO with an $E_{max}\approx350$ TeV \citep{Cao21} and one of the possible TeV counterpart is HESS J1849-000/2HWC J1849+001 \citep{HESS18_GPS, Abeysekara17_2HWC}.

HESS J1849-000 is firmly associated with the PWN G32.6+0.5 generated by the young and energetic PSR J1849-001 \citep{TevCat08, VleeschowerCalas18_J1849} and is one of the the nine sources detected above 56 TeV by HAWC \citep{Abeysekara2020_56TeV}. According \cite{Cao21}, the extension of the detected emission could be coincident also with a YMSC in that region, W43, one of the closest and most luminous star-forming region in the Galaxy hosting a GMC and the Wolf-Rayet WR 121a \citep{TevCat08, Zhang14_W43, Mitchell22_ICRC}. The complexity of the region and several observational challenges cannot give the confirmation of this association \citep{TevCat08} but a study of LHAASO J1849-0003 in the context of other PeV emission detection from similar YMSCs, could indicate if W43 respect all the criteria to explain a UHE emission.  

 No neutrino contraints are present so far for this source.

%Nessuna figura
%................
\paragraph{LHAASO J1908+621}
\label{Sec:J1908}
%................
%Albert22, De Sarkar 22, keyao22, Zhe22, Aharonian HONEST --> modello adronico

\begin{figure}[!h]
    \centering
    \includegraphics[width=0.47\textwidth]{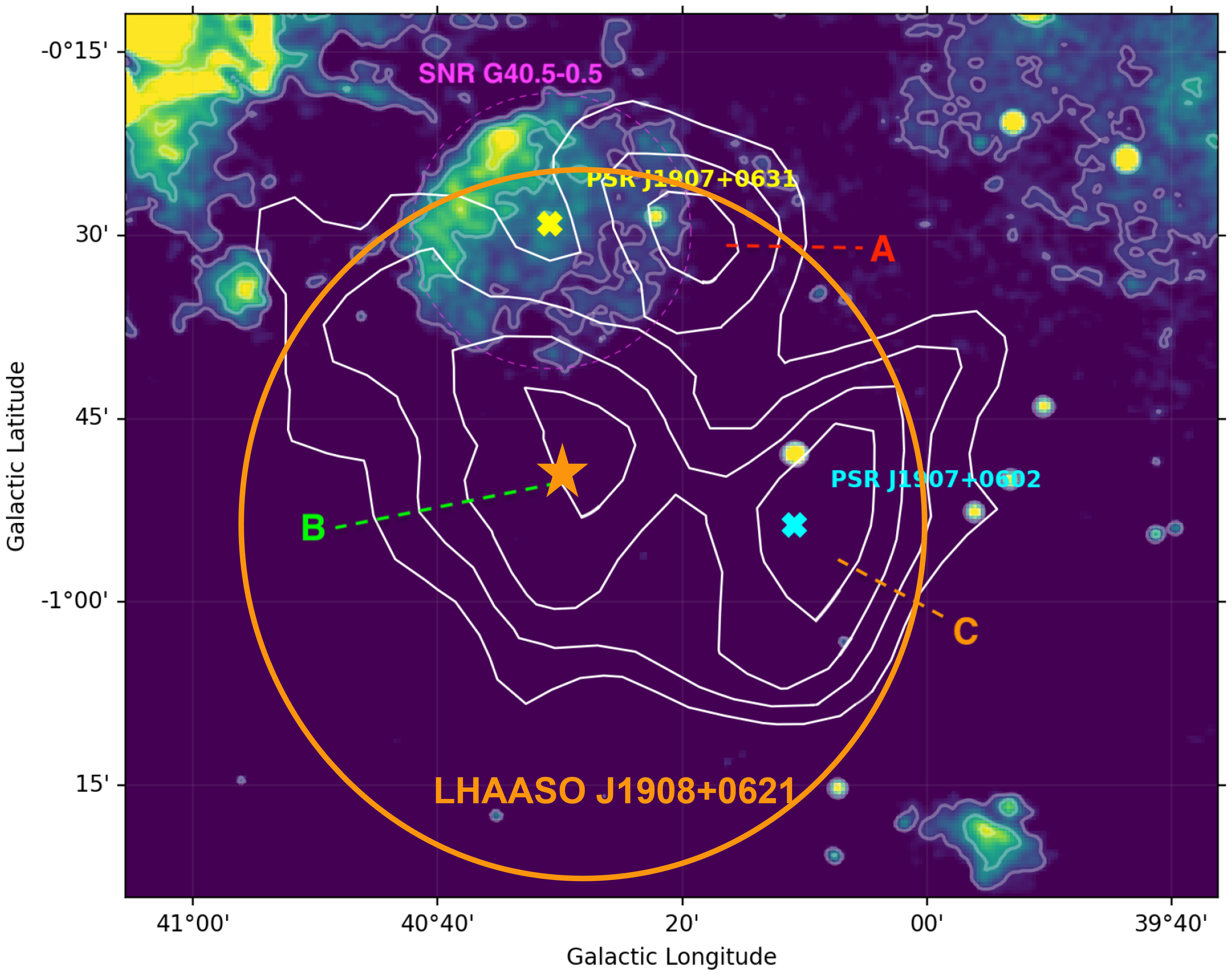}
    \includegraphics[width=0.5\textwidth]{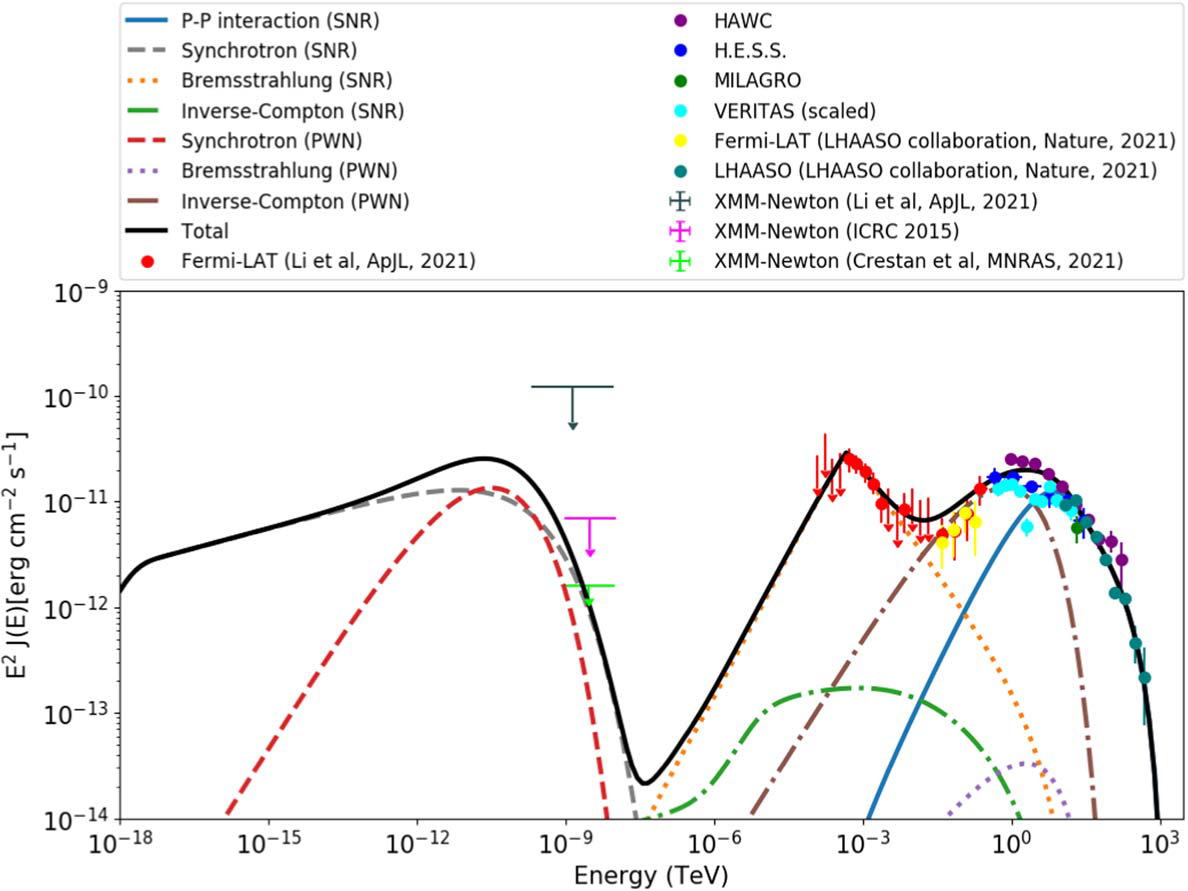}
    \caption{LHAASO J1908+621. Left: the radio map of the LHAASO J1908+621 region from \cite{Crestan21_J1908} with the addition of LHAASO centroid and extension, roughly estimated from the map in \cite{Cao21}. The white solid lines are the VERITAS contours. Right: 
 one of the lepto-hadronic models developed in order to explain the SED of this source where the contribution of hadrons at the highest energies is evident [Figure form \cite{DeSarkar22_J1908}].}
    \label{Fig:J1908}
\end{figure}

LHAASO J1908+621 is one of the LHAASO sources (the other are the Crab Nebula, LHAASO J1825-1326 and LHAASO J2226+6057) detected with the highest significance at E>100 TeV (17.2 $\sigma$) and for which there is a SED at UHE \cite{Cao21}. Its maximum energy is $E_{M}\approx 450$ TeV and shows a spectral steepening maybe due to photon-photon absorption. The potential TeV counterpart is the well known MGRO J1908+06 \citep{Abdo07_MILAGRO_survey} detected also by HESS \citep{Aharonian09_J1908}, ARGO \citep{Bartoli09_J1908}, VERITAS \citep{Aliu14_J1908} and finally by HAWC at E>56 TeV \citep{Albert20_3HWC, Abeysekara2020_56TeV}, with an extension confirmed up to the highest energies. With an off-pulsed analysis, Fermi-LAT detected two GeV excesses in correspondence of the LHAASO J1908+621 contours, Fermi J1906+0626 and 4FGL J1906.2+0631 \citep{Abdollahi20_4FGL}.  
Possible associations are: the middle-aged shell-type SNR G40.5-0.5, that has a MC complex in the surrounding \citep{Li21_1908, Crestan21_J1908}, the PWN of to the radio PSR J1907+0631, and finally the most powerful PSR J1907+0602 (spin-down luminosity sufficient to explain the VHE emission), all in the same region of LHAASO J1908+621 \citep{TevCat08} (see Fig.\ref{Fig:J1908}, left).  

In \cite{Crestan21_J1908}, the authors try to model the whole HE spectrum, concluding that a leptonic or lepto-hadronic models are the most likely with respect to a pure hadronic one. These conclusions are confirmed also by \cite{Albert20_J1908}, with a dominant hadronic component at the highest energies, and then by \cite{DeSarkar22_J1908}, with a leptonic component from the PWN J1907+0602 explaining the GeV-TeV component, and a hadronic component from the SNR/MC interaction for the UHE spectrum (see Fig.\ref{Fig:J1908}, right). The one-component leptonic model is underdog also in the time-dependent model developed in \citep{Joshi23}. Only in \cite{Wu23_J1908} a pure leptonic model is favored but this could depend on the fact that their Fermi-LAT data analysis obtain GeV spectral points different from the ones used in the other papers \citep{Li21_1908}. A cross-check in the GeV band is needed in order to understand better this complex region.

In the IceCube neutrino analysis carried on by \cite{Huang22_neutrini_a}, LHAASO J1908+621 is constrained to be leptonic dominant up to 200 TeV. The bayesian approach, instead, cannot constrain this source, stressing how these estimation are strongly biased by statistical methods \citep{Huang22_neutrini_b}. In \cite{Sarmah23_neutrini}, the authors estimated neutrino fluxes expected in case LHAASO J1908+621 was a SNR with a completely hadronic emission, concluding that the IceCube sensitivity is not enough to detect possible neutrino emission. However, a low-significance neutrino emission seems to come from this region \citep{Aartsen20_IceCube},  supporting the nature of (hadronic) PeVatron of this source.

The higher sensitivity of future neutrino experiments, together with the higher angular resolution of future IACTs, will give a fundamental burst to the understanding of this source. This source will be observable by the ASTRI Mini-Array site for a total of about 550 hours per year and simulations showed that already with 100 hours a high significance spectrum up to 300 TeV can be obtained, perfectly correlated with the one detected by LHAASO \citep{Vercellone22}. Moreover, the ASTRI Mini-Array angular resolution will be able to resolve the possible counterparts in the \gray{} error box \citep{Cardillo22H}. 

%!!!!!!!!!!!!!!!! 
%!!!!!!!!!!!!!!!! 

\begin{figure}[!h]
    \centering
    \includegraphics[width=0.5\textwidth]{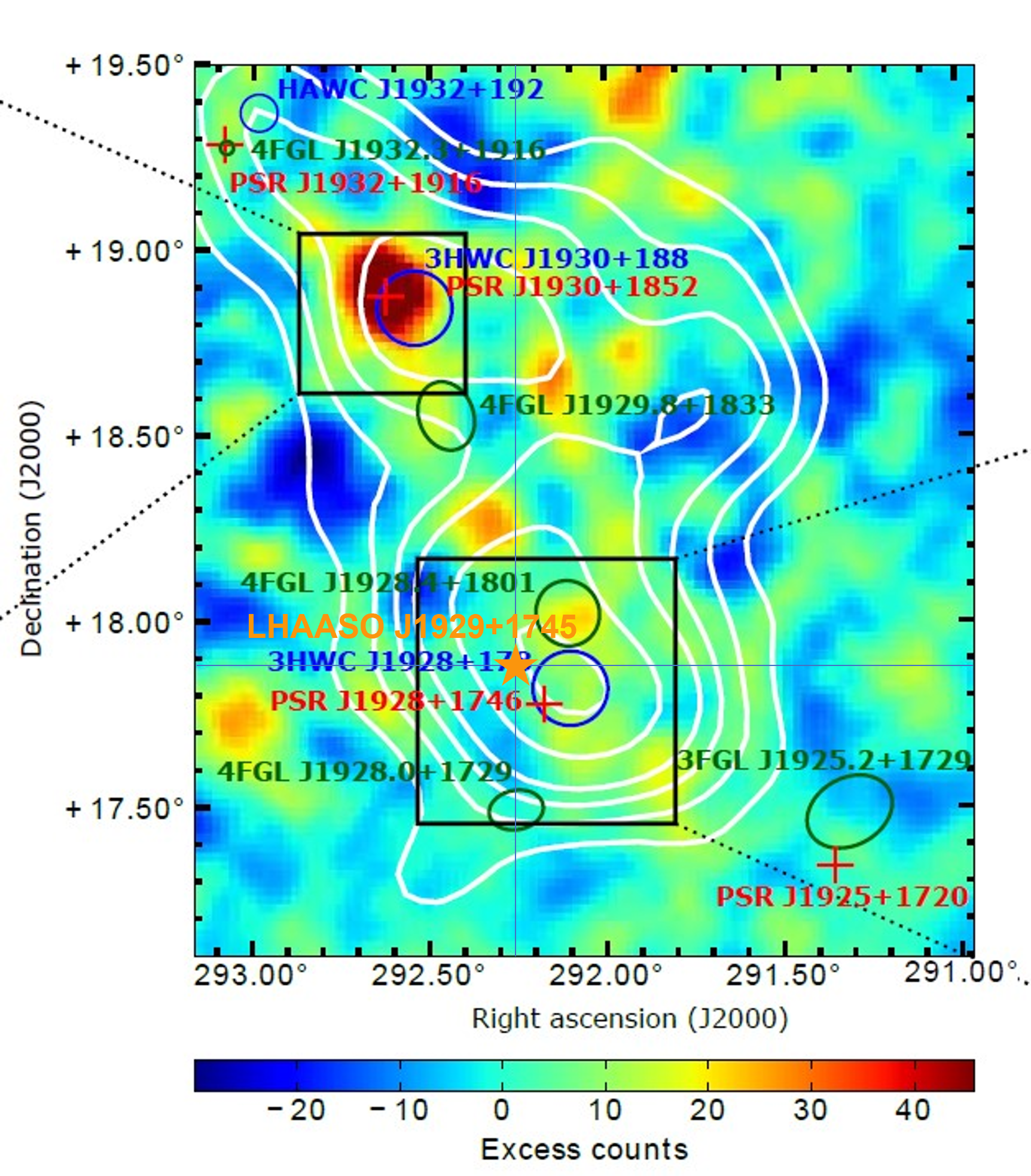}
    \caption{LHAASO J1929+1745. [Figure from \cite{Albert23_J1928_G54.1}] The complexity of the region is well shown in the VERITAS excess map. White contours are HAWC significance contours and the centroid of the LHAASO source is added (we have no hints on its extension).}
    \label{Fig:J1928}
\end{figure}

%................
\paragraph{LHAASO J1929+1745}
\label{Sec:J1829}
%................
%Rbeysekara18 %Albert 2023 Zhang21-22(ICRC)
The maximum energy of LHAASO J1929+1745 is estimated to be $E_{M}\approx700$ TeV \citep{Cao21,TevCat08} and could be associated to two different TeV counterparts: 2HWC J1928+177/HESS J1928+181 and 2HWC J1930+188/HESS J1930+188/VER J1930+188, separated of about 1 degree (see Fig.\ref{Fig:J1928}).

2HWC J1928+177/HESS J1928+181 has almost the same centroid of the LHAASO source and is an extended \gray{} emitter discovered by HAWC \citep{Abeysekara17_2HWC} and then confirmed by HESS \citep{Abdalla21_HAWC_HESS} (VERITAS had no detection because of the large extension of the source \citep{Abeysekara18_VERITAS}). The very recent study of \cite{Albert23_J1928_G54.1} hypothesizes three different origins for the VHE/UHE detected emission, confirmed in the third catalog as 3HWC J1928+178. Since the spatial correspondence with the young PSR 1928+1746, located at 0.03$^{\circ}$, the two most probable ones (the same assumed by \cite{Abdalla21_HAWC_HESS}) are IC emission due to electrons diffuse away from the PWN and CR protons produced by the pulsar then interacting with a nearby MC. 
Actually, the estimated age of the PSR ($\sim 80000$ yrs) and the lack of X-ray detection in its correspondence open the chance that 3HWC J1928+178 could be another TeV halo. This possibility is also favored by the need to add a further extended component to spatially model the surroundings of this source \citep{Albert23_J1928_G54.1}. A recent study \citep{Zhang21_teVhalo} analyzes the affection of the PSR proper motion on the TeV halo morphology, estimating a formula for the maximum possible offset depending on \gray{} energies, PSR transversal velocity and the distance. They concluded that for E>10 TeV an off-set between a pulsar and an associated extended emission cannot be explained by PSR proper motion. In particular, in order to explain LHAASO J1929+1745 with a TeV halo from the PSR 1928+1746, we need a very high transversal velocity ($>2700$ km/s).

The second source, 2HWC J1930+188/HESS J1930+188/VER J1930+188, was detected by all the current IACTs as a point-like source \citep{Acciari10_G54.1, Abeysekara17_2HWC,HESS18_GPS} and it is strongly associated with the PWN in the 2000 yrs old SNR G54.1+0.3 \citep{TevCat08}, with no chance to understand if the \gray{} emission is from the shell or from the PWN. This has also a point-like GeV no-pulsed counterpart detected by Fermi-LAT, 3FGL J1928.9+1739 \citep{Abeysekara18_VERITAS,TevCat08}. The last analysis by \cite{Albert23_J1928_G54.1} confirmed the VHE steep spectrum detected by VERITAS that could be a hint of the presence of a cut-off (to faint to be validated). If this cut-off will be confirmed, this source could be ruled out as the origin of the UHE \gray{} emission detected by LHAASO.

 No neutrino constraints are present for this source so far.

%!!!!!!!!!!!!!!!! 
%!!!!!!!!!!!!!!!! 

%................
\paragraph{LHAASO J1956+2845}
\label{Sec:J1956}
%................
%TeVCat, Xing22

\begin{figure}[!h]
    \centering
    \includegraphics[width=0.6\textwidth]{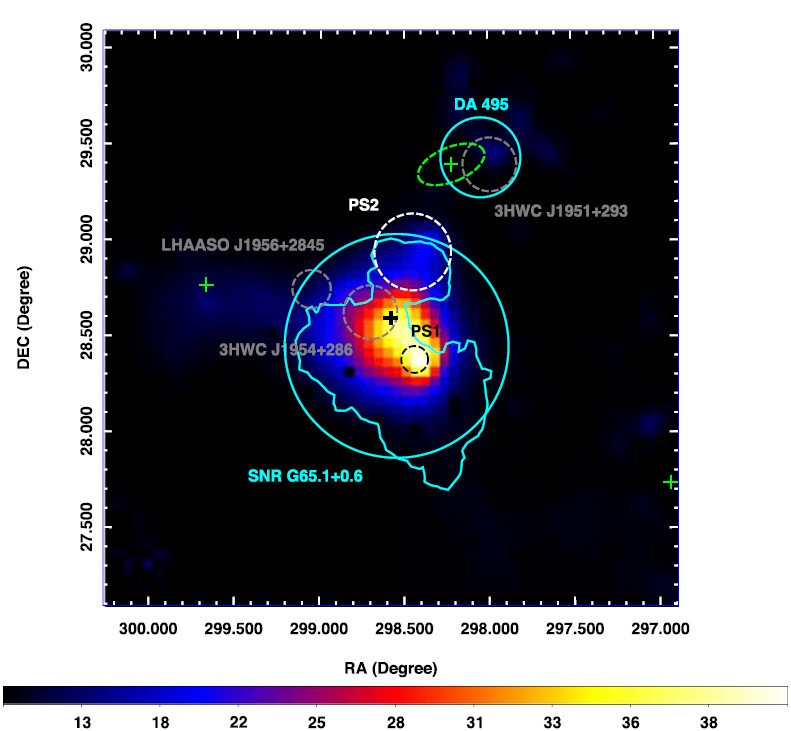}
    \caption{LHAASO J1956+2845. [Figure from \cite{Xing22_J1956}] The Fermi-LAT off-pulse TS map for the region of PSR J1954+2836 for 0.5-500 GeV, where all the catalogue sources are removed. The cyan contour is an approximate radio shape of the SNR G65.1+0.6.}
    \label{Fig:J1956}
\end{figure}

LHAASO J1956+2845, with an $E_{M}\approx 420$ TeV \citep{Cao21}, has just one TeV potential counterpart, the 2HWC J1955+285/3HWC J1954+286 \citep{TevCat08}, even if it is $ 0.33^\circ$ away (see Fig. \ref{Fig:J1956}). Two candidate sources exist for the origin of VHE/UHE emission: the shell-type middle aged radio SNR G65.1+0.6 and the pulsar PSR J1954+2836 detected by Fermi-LAT at $E>10$ GeV \citep{TevCat08, Xing22_J1956}.

The analysis of the GeV \gray{} data from the Fermi-LAT is the first (and the only so far) attempt to disentangled SNR and PSR emission in order to associate the VHE/UHE emission to one of the sources \citep{Xing22_J1956}. In this work, the GeV emission is attributed to SNR G65.1+0.6 that is unlikely to be the source of the VHE emission because of the no-spatial coincidence between the HAWC and Fermi-LAT emission. On the other hand, no pulsating emission has been detected from the PSR J1954+2836; however, the authors estimated a possible TeV halo generated from this source (based on Geminga one parameter), suggesting that this is the VHE/UHE emission source. 

One more time, it is evident the need to resolve VHE/UHE emission region morphology with high spatial resolution analysis and to carry on parallel study with a constant comparison with the other known TeV Halos and/or PWN in order to analyse the contraints to be Galactic CR accelerators.

 No neutrino constraints are present for this source so far.

%!!!!!!!!!!!!!!!! 
%!!!!!!!!!!!!!!!! 

%................
\paragraph{LHAASO J2018+3651}
\label{Sec:J2018}
%................
%Yang23, HAWC21, Aliu14

LHAASO J2018+3651 has a maximum energy of about 270 TeV and it is one of the three LHAASO UHE detections located in the Cygnus region (the others are LHAASO J2032+4102 and LHAASO J2108+5157), and has two potential TeV candidates: VER J2016+371 and VER J2019+368 \citep{Cao21,TevCat08}.

\begin{figure}[!h]
    \centering
    \includegraphics[width=0.45\textwidth]{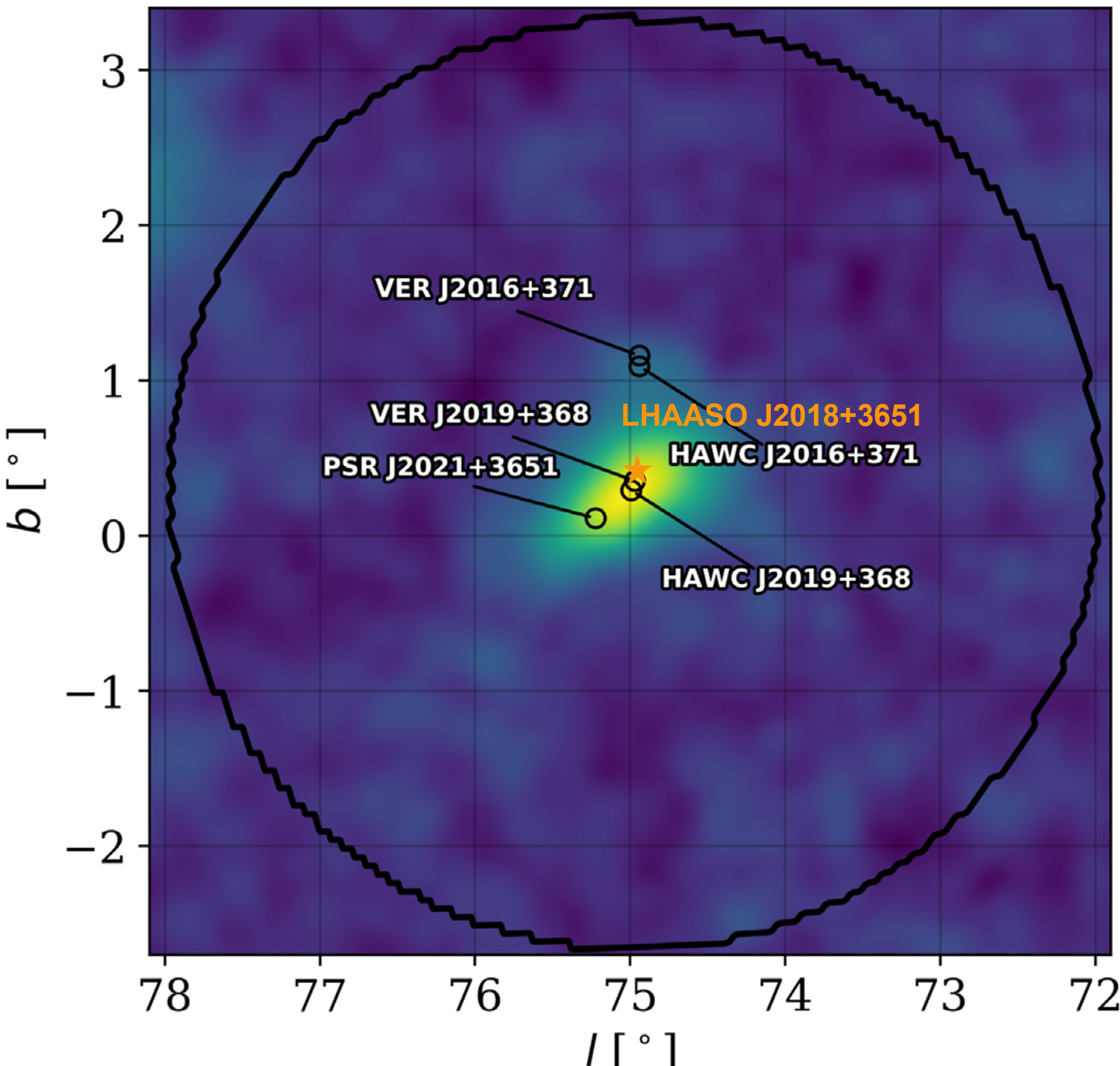}
    \includegraphics[width=0.53\textwidth]{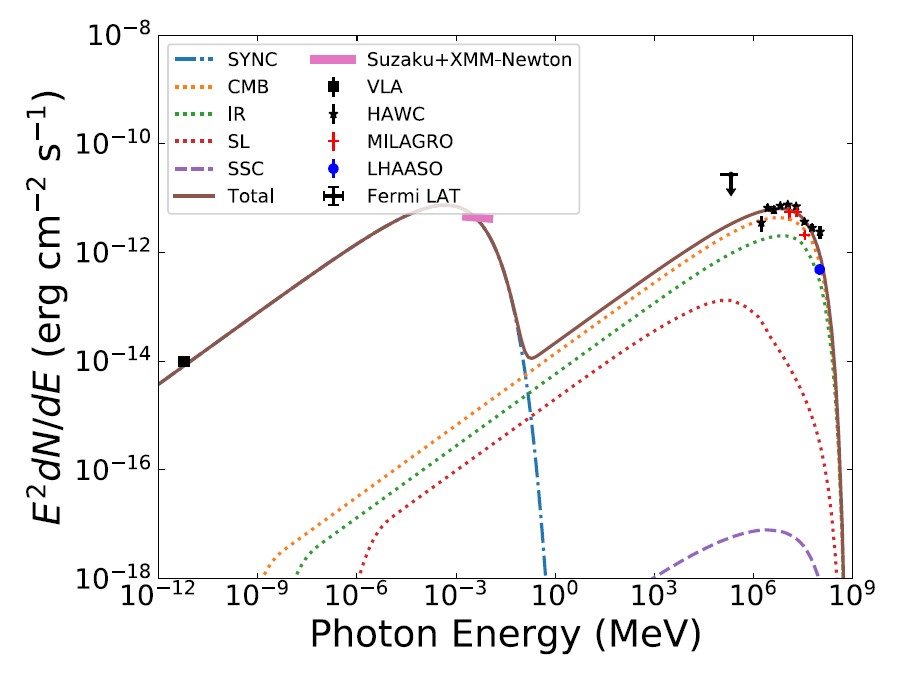}
    \caption{LHAASO J2018+3651. Left: the HAWC significance map of the region of interestwith the addition of the LHAASO centroid [Figure from \cite{Albert21_J2019}]. Right: the SED of the PWN G75.2+0.1 where the leptonic model can perfectly explain the multi-wavelength spectrum from this source [Figure from \cite{Yang23_J2019}].}
    \label{Fig:J2018}
\end{figure}

These two VERITAS sources are both associated with MGRO J2019+37, an extended \gray{} emission detected the first time by the Milagro observatory \citep{Abdo07_J2019}. This region, one of the brightest of the Cygnus region, overlaps a lot of different sources but the PWN of the PSR J2021+3651 together with the star-forming region Sharpless 104 (Sh 2-104) were supposed to be the main contributors \citep{Saha15_J2019,Paredes09_J2019}  (see Fig.\ref{Fig:J2018}, left). Later, VERITAS resolved this TeV region in the two different sources candidates as LHAASO J2018+3651 counterpart, even if the extension of the LHAASO emission cannot be resolved with the same angular resolution.

The first candidate, VER J2016+371, is an unidentified point-like source spatially coincident with part of MGRO J2019+37 \citep{Aliu14_J2019} and confirmed also by HAWC (HAWC J2016+371) \citep{Albert21_J2019}. The most likely counterpart is the PWN in the SNR CTB 87 because of both the co-location of VHE and X-ray emission and its luminosity in the two bands \citep{Albert21_J2019}.

The other candidate, VER J2019+368, is a bright extended source (1 degree), with an energy dependent morphology, and represents the bulk of the emission from  MGRO J2019+37. Its emission was confirmed also by HAWC (HAWC J2019+368) \citep{Albert21_J2019} with a high integrated flux above 56 TeV. The morphological and spectral features in both TeV and X-ray bands points toward a PWN origin for this VHE/UHE emission \citep{Albert21_J2019}, precisely from the Dragonfly PWN (G75.2+0.1) powered by the PSR J2021+3561. This hypothesis is supported by a very recent multi-wavelength work from radio to X-ray to UHE \gray{}, taking into account also the LHAASO spectral point \citep{Yang23_J2019} (see Fig.\ref{Fig:J2018}, right).

Even if we cannot say if LHAASO J2018+3651 UHE detection is in correspondence on VER J2016+371 or VER J2019+368, according to the models and considerations developed up to now, the contribution from the star forming region Sh 2-104 seems to be not the dominant one.

 No neutrino constraints are present so far for this source.

%!!!!!!!!!!!!!!!! 
%!!!!!!!!!!!!!!!! 
%................
\paragraph{LHAASO J2032+4102}
\label{Sec:Cocoon}
%................
%Cugnus Cocoon --> Aharonian 19 e Aharonian HONEST, Banik 22, Bykov+22, guevel22, DeOnaWilhelmi22

\begin{figure}[!h]
    \centering
    \includegraphics[width=0.5\textwidth]{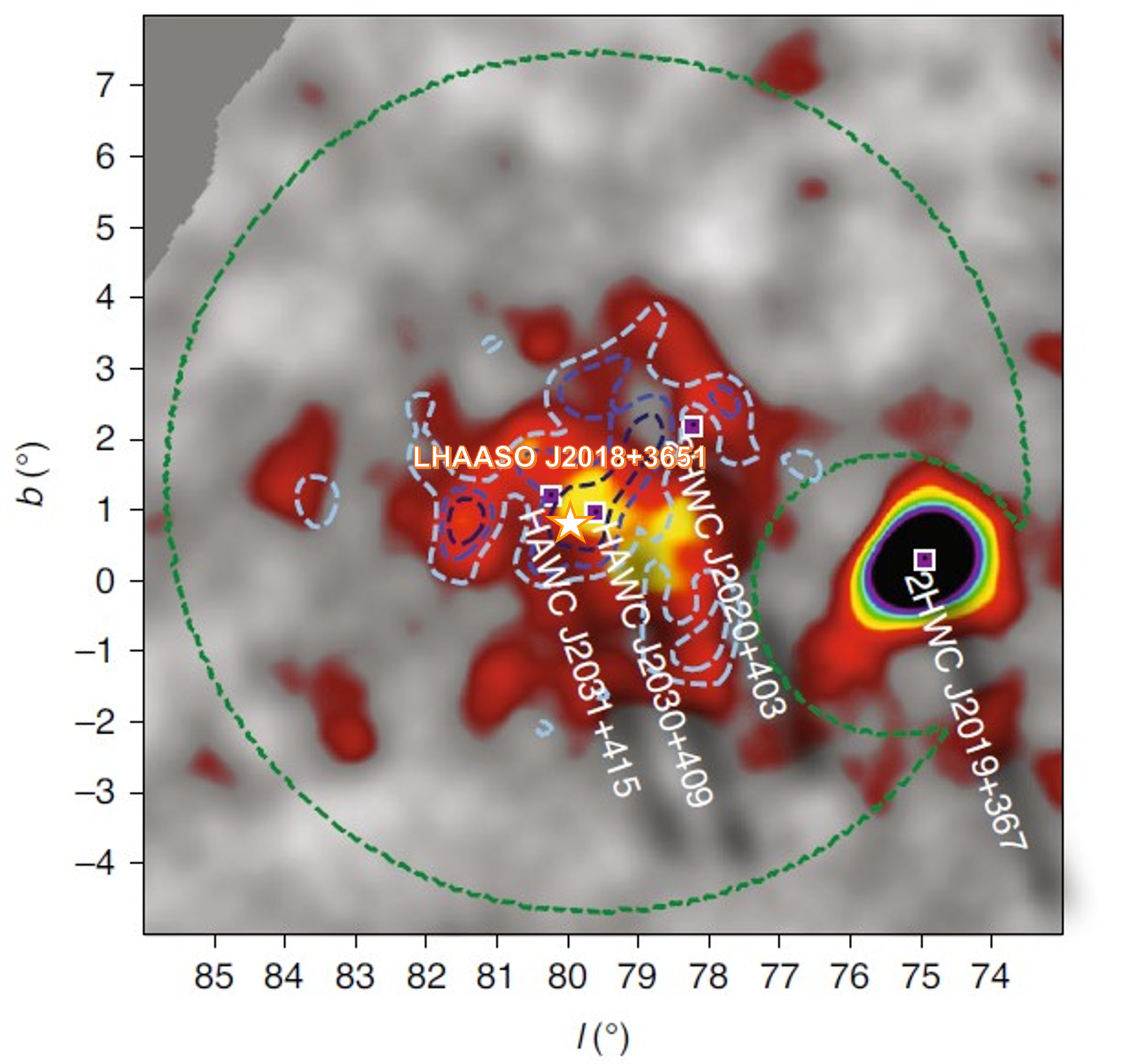}
    \includegraphics[width=0.49\textwidth]{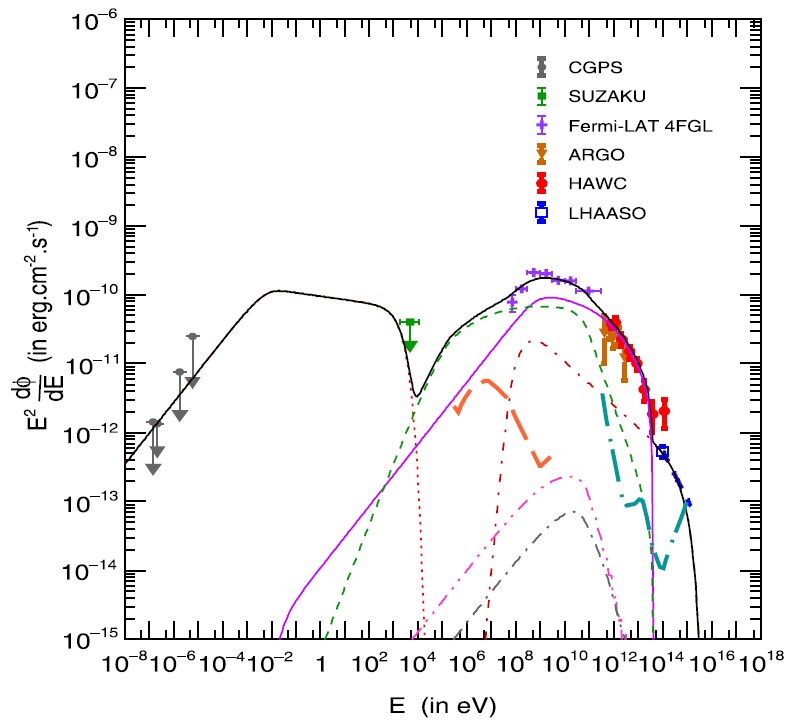}
    \caption{LHAASO J2032+4102. Left: HAWC significance map of the Cocoon Region after the subtraction of the known sources, with the addition of the LHAASO source centroid [Figure from \cite{Abeysekara21_Cocoon}]. Right: the multi-wavelength data from the Cygnus cocoon fitted by the lepto-hadronic scenario in \cite{Banik22_cocoon} [Figure from in \cite{Banik22_cocoon}]. From this model the authors estimated the neutrino flux expected if the source is hadronic .}
    \label{Fig:Cocoon}
\end{figure}

LHAASO J2032+4102 has the highest maximum energy between the 12 PeVatrons candidates: 1.4 PeV. As like as LHAASO J2018+3651 and (the next) LHAASO J2108+5157, this is located in the Cygnus region and the potential TeV counterparts are several candidates:  TeV J2032+4130/2HWC J2031+415/VER J2032+414, ARGO J2031+4157 and MGRO J2031+41 (see Fig.\ref{Fig:Cocoon}). 

The first source, TeV J2032+4130/2HWC J2031+415/VER J2032+414, was detected at TeV energies firstly by HEGRA \citep{Aharonian02_J2032} and then by HAWC \citep{Abeysekara17_2HWC} and VERITAS \citep{MAGIC_VERITAS18_J2032}. It is associated with the PWN of the PSR J2032+412 and overlaps with one of the nine sources of the first HAWC catalog emitting at E>56 TeV \citep{TevCat08}. According to \cite{Linden17_TeV_halos}, it is a candidate TeV halo because it is associated with a pulsar >100 kyrs and has a TeV halo-expected-flux large at least 2$\%$ of the Geminga flux. However, no other analysis has confirmed (or ruled out) this hypothesis. According the graph produced in the work \cite{DeOnaWilhelmi22} (see Fig. \ref{Fig:PWNe}), LHAASO J2032+4102 is one of the two (the other is LHAASO J2108+5157 that has any PSR in the surroundings) LHAASO candidate Pevatrons that cannot be explained by an associated pulsar; it lies above the absolute maximum electron energy producible by a PSR limited by radiative losses or potential drop (see Sect.\ref{Sec:PWNe}). This important work indicates that we need to find other sources as the origin of the UHE LHAASO detected emission from this region.

In this context, an important candidate is ARGO J2031+4157, that is the TeV counterpart of the Cygnuns Cocoon \citep{TevCat08}, detected for the first time at GeV energies by Fermi-LAT \citep{Ackermann11_Cocoon} and then confirmed at E>100 TeV by HAWC \citep{Abeysekara21_Cocoon} and Tibet AS$\gamma$ \citep{Amenomori21_Cygnus}. In \cite{Bykov22_cocoon} the authors try to explain the whole GeV-TeV spectrum of the Cygnus Cocoon with a model based on \cite{Bykov01}: particle acceleration and propagation in a superbubble with multiple shocks of different strengths produced by powerful winds of massive stars and supernovae. Their model well explains Fermi-LAT, ARGO and HAWC data but it doesn't fit the only one LHAASO spectral point. However, it can explain also the behavior of the \gray{} spectrum from Westerlund 2, suggesting that this model can work in MSCs and questioning the association of the LHAASO spectral point with Cygnus Cocoon. Further and deeper analysis is needed in order to confirm this conclusion.

In \cite{Banik22_cocoon}, with a simpler acceleration model without assumptions on propagation, the authors explain the multi-wavelength spectrum of the Cygnus region with a lepto-hadronic scenario (see Fig.\ref{Fig:Cocoon}, right), estimating also a neutrino flux in agreement with an IceCube neutrino (see Fig.4 of \cite{Banik22_cocoon}, right).
The Cygnus Cocoon, indeed, could be the only Galactic source with the evidence of neutrino emission (IceCube-201120A, at the IceCube sensitivity limit, not confirmed), coincident with a \gray{} excess detected by Carpet-2 experiment \citep{Dzhappuev21_cocoon_neutrino}. If this neutrino event was correlated with the Cocoon, taking into account also the UHE LHAASO photons, this could be a hint that this source could be not only a confirmed LHAASO PeVatron but a hadronic one. 

The last possible TeV counterpart, MGRO J2031+41, could be identified with TeV J2032+4130 but Milagro speculated that its emission could due to more than one source and, consequently, it is listed separately in the TeVCat \citep{TevCat08}.

%!!!!!!!!!!!!!!!! 
%!!!!!!!!!!!!!!!! 

%................
\paragraph{LHAASO J2108+5157}
\label{Sec:J2108}
%................
%Abe+22,DeLaFuente23,DeonaWilhelmi 2022, Cao21_J2108

\begin{figure}[!h]
    \centering
    \includegraphics[width=0.5\textwidth]{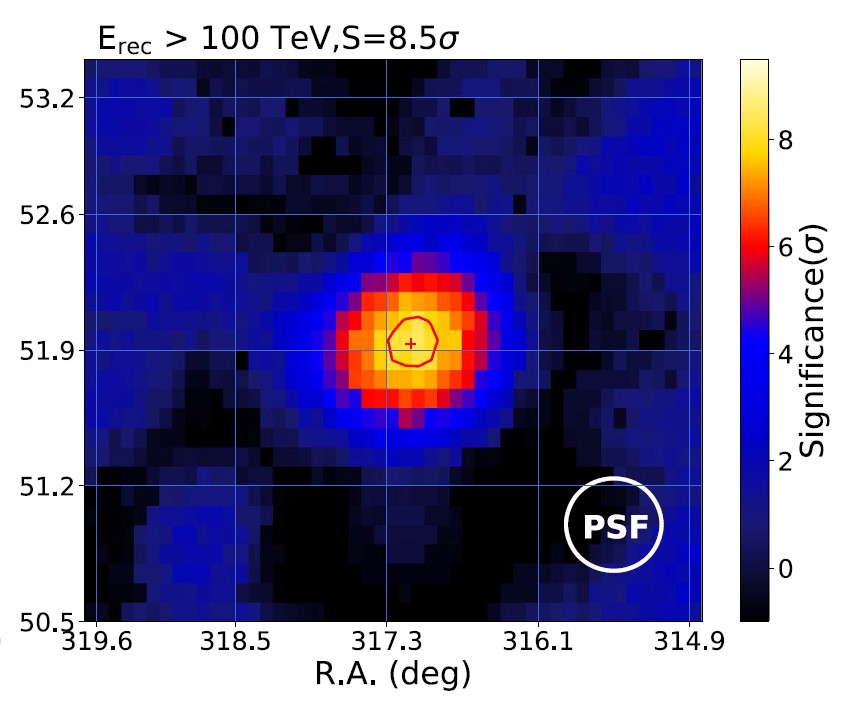}
    \includegraphics[width=0.48\textwidth]{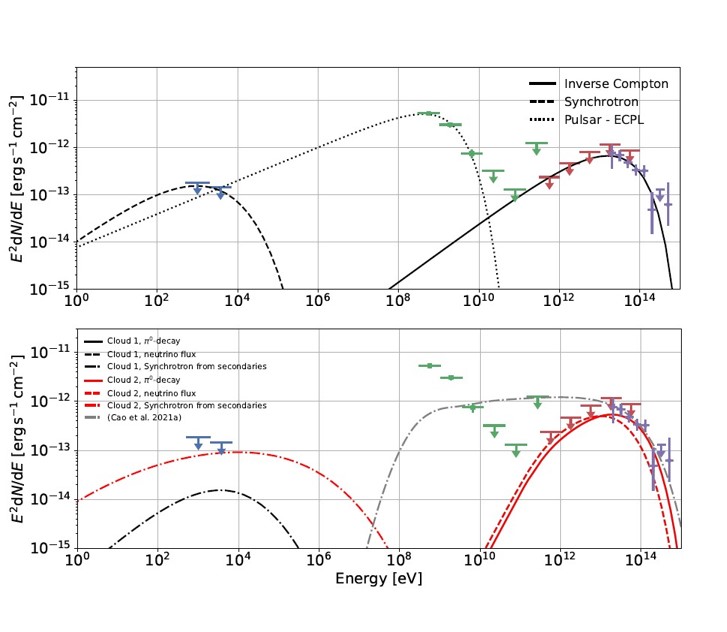}
    \caption{LHAASO J2108+5157. Left: significance map around LHAASO J2108+5157 as observed by KM2A at E>100 TeV [Figure from \cite{CAo21_J2108}]. Right: the leptonic (top) and hadronic (bottom) that can explain the multi-wavelength SED of this source in \cite{Abe22_J2108} [Figure from \cite{Abe22_J2108}].}
    \label{Fig:J2108}
\end{figure}

LHAASO J2108+5157 is a unique case because has no detected TeV counterparts and it is the first source ever detected for the first time at UHE \citep{CAo21_J2108,TevCat08} (see Fig.\ref{Fig:J2108}, left). Its maximum energy is about 430 TeV and is found to be point-like without, however, the chance to rule out a possible extension. This is located in the complex Cygnus region (CygnusOB7 MC) and it is the only source without a bright pulsar found in the surroundings (the first is at 3$^{\circ}$)\citep{DeOnaWilhelmi22}.
Fermi-LAT extended source 4FGL J2108+5155e was found in its proximity ($\sim0.13^\circ$) but a physical connection is not obvious because of spectral differences \citep{CAo21_J2108}. 

Instead, the correlation with the MC [MML2017]4607 seems to be confirmed \citep{CAo21_J2108} and a possible second cloud is proposed in \cite{DeLaFuente23_J2108} where the authors present a pioneering study of the COB7 region with low-resolution $^{12,13}$CO(2$\rightarrow$1) observations made with the 1.85m radio-telescope at Osaka Prefecture University in order to investigate the nature of the LHAASO sources. This second cloud is [FKT-MC]2022 and the authors proposed it as candidate source of the UHE from LHAASO J2108+5257, with a favoured hadronic origin (MCs illuminated by accelerated CRs). Consequently, we have two MCs that encourage a hadronic scenario \citep[see also][]{Kar22_J2108}, even if the hard spectral index requires non-linear DSA or stochastic acceleration hypothesis. 

The multi-wavelength analysis carried on in \cite{Abe22_J2108} putting together LST-1, XMM-Newton and Fermi-LAT results, cannot rule out any scenario, leptonic or hadronic (see Fig.\ref{Fig:J2108}, right). LST-1 and LHAASO data can be explained by IC scenario but the absence of X-ray emission implies a weak magnetic field, favoring a TeV Halo scenario over that the PWN one. However, there is no hint of a pulsar. Anyway, in their work a more complete combined analysis of both leptonic and hadronic components in the region is missing and this could obtain more constraining parameters.

The estimated neutrino flux in a complete hadronic scenario is in the sensitivity range of future instruments, giving to us a chance to understand this LHAASO challenging source \citep{Kar22_J2108}.

%!!!!!!!!!!!!!!!! 
%!!!!!!!!!!!!!!!! 

%................
\paragraph{LHAASO J2226+6057}
\label{Sec:Boomerang}
%................
%Boomerang SNR --> Safi HONEST, Ge+21, fang+22, de sarkar22,  Liang22, Liu22, Liu20, MAGIC22, Verna22

LHAASO J2226+6057 is one of the four LHAASO sources (the other are the Crab Nebula, LHAASO J1825-1326 and LHAASO J1908+621) detected with the highest significance at E>100 TeV (13.6 $\sigma$) and for which there is a SED at UHE. Its maximum energy is $E_{M}\approx570$ TeV and shows a spectral steepening maybe due to photon-photon absorption \cite{TevCat08,Cao21}. Its TeV counterparts is VER J2227+608/HAWC J2227+610 \citep{TevCat08} and there are two possible sources explaining this VHE/UHE emission: the SNR G106.3+2.7, with the associated MC, and the Boomerang PWN, associated to the PSR J2229+6141. The SNR is located in the "tail" of the VHE emission and the PWN on the "head".

\begin{figure}[!ht]
    \centering
    \includegraphics[width=0.95\textwidth]{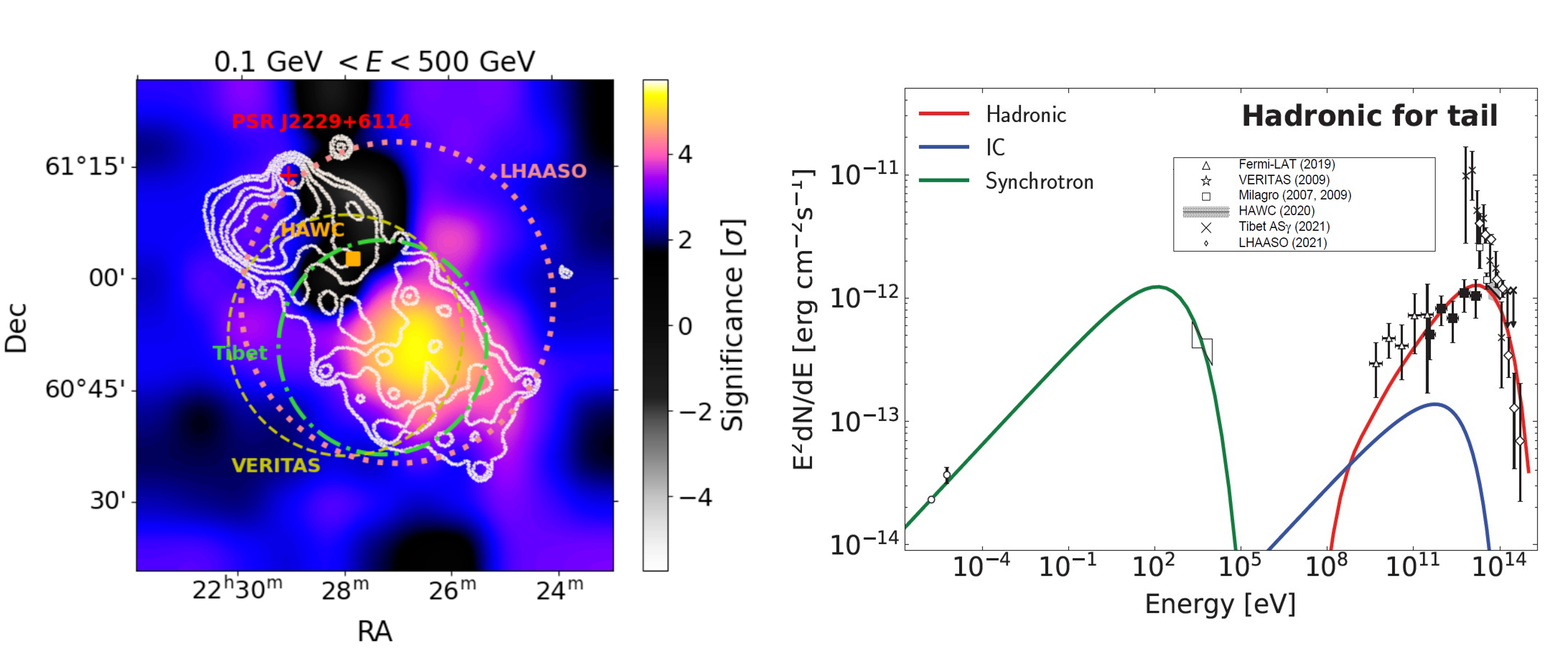}
    \includegraphics[width=0.48\textwidth]{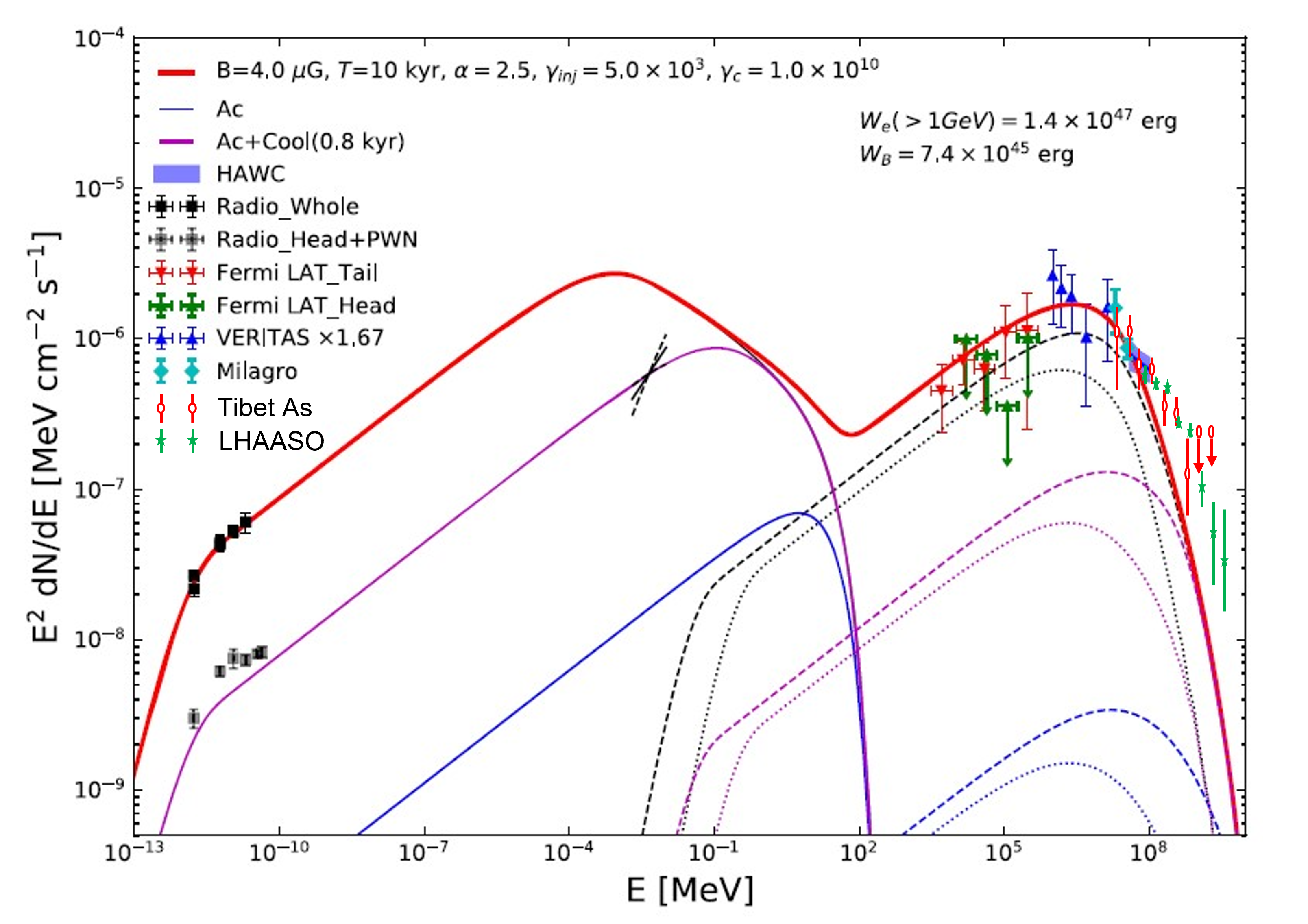}
    \includegraphics[width=0.48\textwidth]{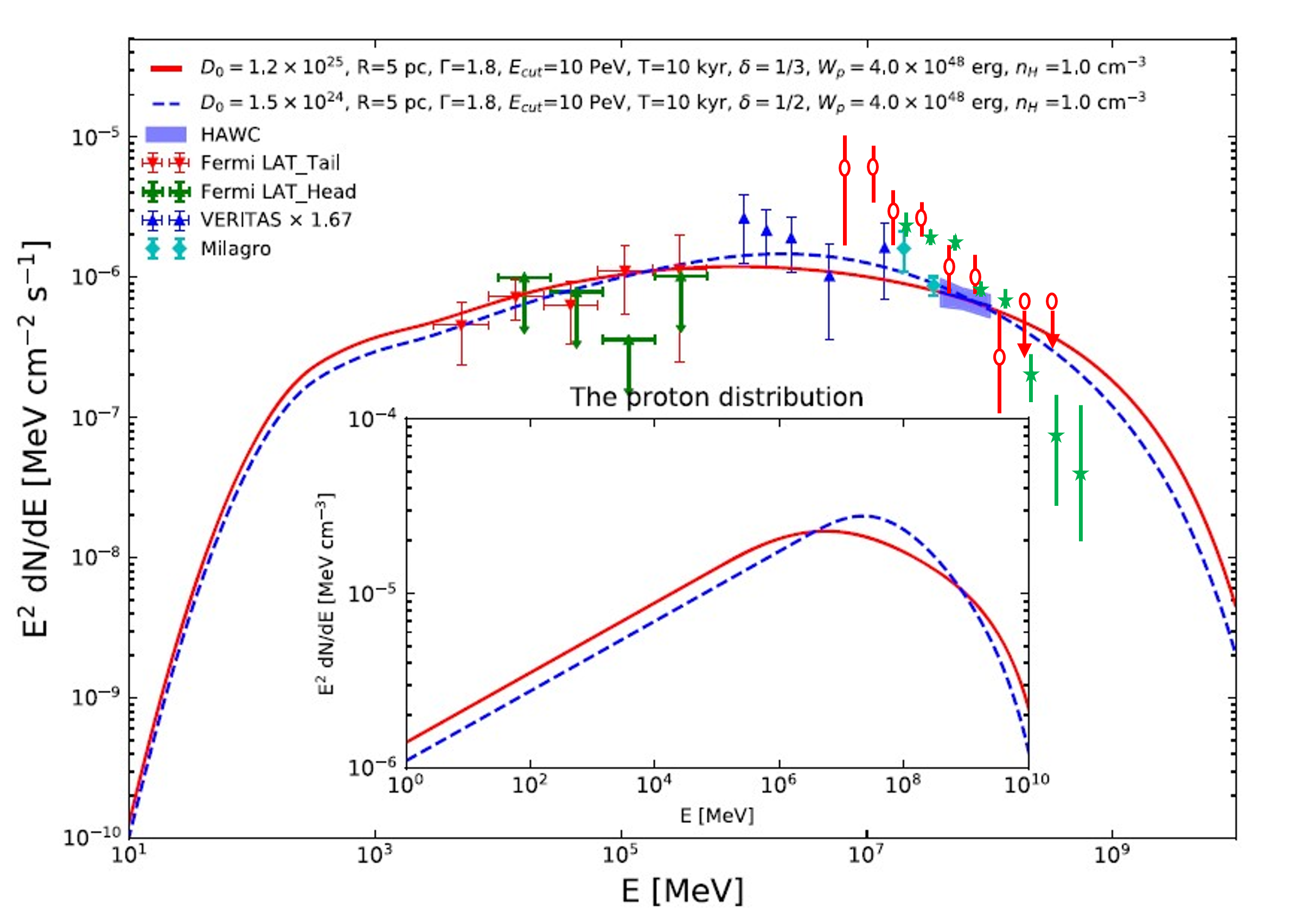}
    \caption{LHAASO J2226+6057. Top left: Fermi-LAT residual significance map of the region for 0.1-500 GeV where is evident the concentration of the GeV emission in the tail region. White contours are the radio continuum emission at 1420 MHz [Figure from from \cite{Fang22_Boomerang} ]. Top right: the hadronic model used by MAGIC collaboration to fit all data from the tail region. Milagro, HAWC, TibetAS$\gamma$ and LHAASO shown here are from extraction regions that partially include the head, consequently they could be contaminated [Figure from \citep{MAGIC23_Boomerang}]. Bottom: Spectral fit to the spectral energy distribution of Ver J2227+608 from \cite{Liu20_Boomerang}, leptonic model on the left and the hadronic one on the right. This was developed before the last Tibet AS$\gamma$ and LHAASO data, that we have overlapped, showing that they rule out both the models [Figure from \citep{Liu20_Boomerang}].}
    \label{Fig:Boomerang}
\end{figure}

\begin{figure}[!ht]
    \centering
    \includegraphics[width=1\textwidth]{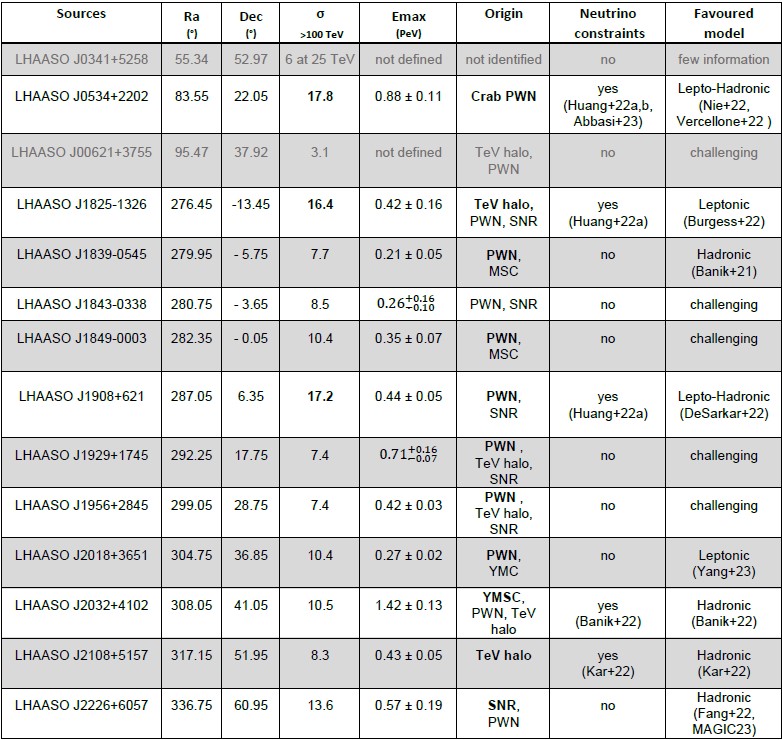}
    \caption{A summary table of the all LHAASO VHE sources with coordinates, significance at 100 TeV, maximum energy from \cite{Cao21}. the last three columns are based on the review in Section \ref{Sec:12PeV}: the "origin" column has in boldface the most probable kind of source, the "neutrino constraints" indicates if we can have information on \gray{} emission from neutrino analysis/estimation and the last column indicated the most likely origin of the emission according the analysis done up to now.}
    \label{Fig:Cardillo_table}
\end{figure}

The UHE detection by HAWC \citep{Albert20_Boomerang}, Tibet AS$\gamma$ \citep{Amenomori21_Cygnus} and finally by LHAASO \citep{Cao21} has a low angular resolution and, consequently, we were not able to say if it is from the head or the tail region. Recently, 12 years Fermi-LAT GeV analysis of the region shows that at the highest energies (10-500 Gev), only the tail is emitting \gray{} (see Fig.\ref{Fig:Boomerang}, top left) and a hadronic model from the SNR/MC interaction can explain the whole HE/VHE/UHE spectrum \citep{Fang22_Boomerang}. Very recently, the MAGIC collaboration resolved the VHE/UHE emission and detected E>10 TeV only from the tail region\citep{MAGIC21_Boomerang, MAGIC23_Boomerang}, leaning towards hadronic explanation of \gray{} emission \citep{Albert20_Boomerang, MAGIC23_Boomerang}, linked to the SNR/MC interaction (or MC illumination by CRs escaping from the SNR \citep{Liu22_Boomerang}). The hadronic model used by the MAGIC collaboration takes into account also the GeV Fermi-LAT data points from the tail region from \cite{Xin19_Boomerang} (consistent with the more recent results by \cite{Fang22_Boomerang}) (see Fig.\ref{Fig:Boomerang}, top right).

The same detection was explained with hadronic emission but from the PWN and not from the SNR by \cite{Xin19_Boomerang}, since the hard spectral index and low energy content of CRs. The PWN origin is investigated also by \cite{Liang22_Boomerang}, concluding that, if a part of the UHE \gray{} emission comes from the PWN, it has to be of hadronic origin.

The real origin of the hadronic emission will be understood only with a deep analysis of the micro-physics of the region but, in spite of PWN or SNR origin, the hadronic origin is supported also by the analysis on non-thermal X-ray radiation \citep{Ge21_Boomerang, Fujita21_Boomerang}, detected everywhere but with an enhancement of the luminosity in the head region, well correlated with the radio emission but no with the \gray{} one.

%!!!!!!!!!!!!!!!! 
%!!!!!!!!!!!!!!!! 

Nevertheless, there are some works that analyse possible leptonic or lepto-hadronic origin of UHE emission of LHAASO J2226+6057. In \cite{Liu20_Boomerang}, before the LHAASO and Tibet AS$\gamma$ detection, the whole spectrum from VER J2227+608 can be explained with both leptonic and hadronic scenario and the authors conclude that only UHE data can discern the two models. However, an overlap of the last LHAASO and Tibet AS$\gamma$ data on both their models makes clear that neither model is confirmed (see Fig. \ref{Fig:Boomerang}, bottom) requiring a further analysis with constraints on parameters computed by the LHAASO results. In both \cite{DeSarkar22_Boomerang} and \cite{Joshi23}, the authors developed a simplistic one-zone leptonic model (but complete of reverberation effect and PWN age estimation) assuming that the UHE emission comes from the Boomerang PWN. They obtain a good fit but challenging low value of the magnetic field and a PWN size too large maybe due to the very simple model used. 

It is clear that this source is very interesting for the future IACTs as ASTRI Mini-Array and CTA. Both the collaboration have made simulations in this region in order to valuate the contribution of these two instruments in its understanding. In \cite{Verna22_Boomerang_CTAN} is shown how the CTA-North could reconstruct the spectrum with a very high precision; in \cite{Vercellone22}, instead, is shown that a detection of the ASTRI Mini-Array could disentangled hadronic and leptonic models. Moreover, the very good angular resolution of this array could resolve definitely the VHE emission location \citep{Cardillo22H}.

Assuming a completely hadronic origin of its \gray{} emission, in \cite{Sarmah23_neutrini} the authors estimated the neutrino fluxes expected concluding that the IceCube sensitivity is not enough to detect possible neutrino emission. 

%!!!!!!!!!!!!!!!! 
%!!!!!!!!!!!!!!!! 

%%%%%%%%%%%%%%%%%%%%%%%%%%%%%%%%%%%%%%%%%%
\section{The Future}
\label{Sec:future}
%%%%%%%%%%%%%%%%%%%%%%%%%%%%%%%%%%%%%%%%%%

All the space and "on-Earth" facilities allowed us to know a lot about CRs and their sources. However, this was not sufficient to have a definitive answer to the question "What is the CR origin?". 

After the review of all the LHAASO candidate PeVatrons it's quite clear what are the ingredients to have a better understanding of Galactic CRs:
\begin{itemize}
    \item Higher angular Resolution of the future VHE/UHE \gray{} instruments;
    \item Higher sensitivity of the future neutrino experiments;
    \item Multi-wavelength and multi-messengers analysis of the same source;
    \item Improvement in the comprehension of the micro-physics of different kind of sources.
\end{itemize}

According to the observational evidence, a very high angular resolution is the most important feature in an instrument to look into the sources. It will allow a much better source association of detected \gray{} emission, a deeper understanding of possible energy dependent morphology and to strengthen multi-wavelength correlation. Highest energy sensitivity is also fundamental but LHAASO has proven that this is sufficient to identify the presence of PeVatrons in our Galaxy but not to have a firm association with a parent source and them to discern hadronic from leptonic emission.
Morphological information will be crucial also to understand some specific parameters of candidate PeVatrons. For example, compact sub-degree regions in PWNe indicate high magnetic field and fast synchrotron losses, and a detection of extended X-ray nebula should mean that LHAASO \gray{} detection comes from the up-scattering of the 2.7 K CMB radiation in PWNe \citep{DeOnaWilhelmi22}. A similar detection is challenging for current X-ray instruments (XMM-Newton, Chandra) but will be possible for the eROSITA satellite that will have a larger FoV \citep{DeOnaWilhelmi22}. 

\begin{figure}[!h]
    \centering  
    \includegraphics[width=0.495\textwidth]{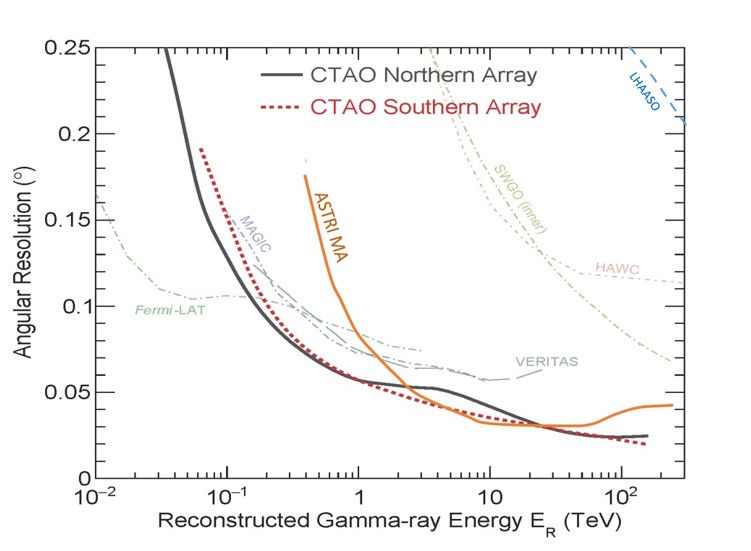}
    \includegraphics[width=0.495\textwidth]{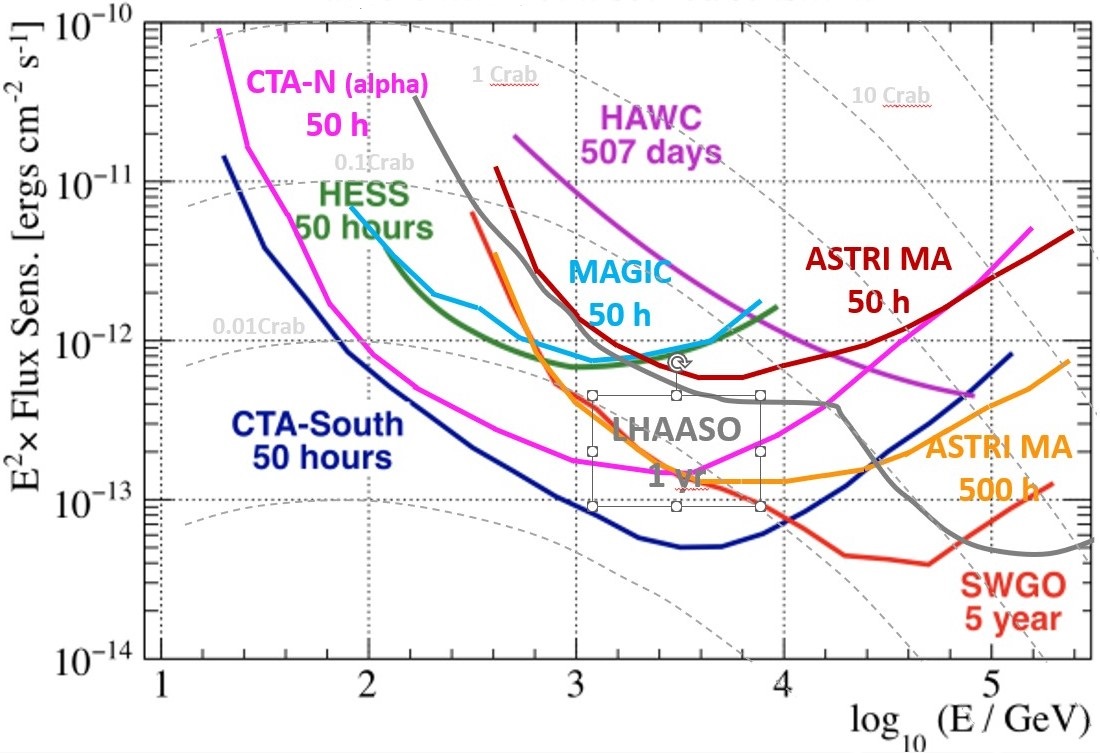}
    \caption{Left: Angular Resolution of current and future instruments from \cite{CTA_performances} with the addition of the ASTRI Mini-Array updated curve with optimized cuts from \citep{Lombardi22} and the LHAASO curve from \cite{Hinton22H}. Right: Differential sensitivity curves of current and future VHE/UHE instruments from \cite{SafiHarb22H}, expanded with the addition of the ASTRI Mini-Array curves for 50h and 500h and other instrument curves \citep{Lombardi22}.}
    \label{Fig:performance_curves}
\end{figure}

Even if the IACTs have some evident disadvantages with respect to the EAS arrays (e.g. limited duty cycle (10-15 $\%$) or limited FoV), their low-energy threshold and high angular resolution ($< 0.1^{\circ}$ at 1 TeV) makes IACTs ideal for detailed morphological studies. For this reasons, instruments as ASTRI Mini-Array and CTA will make the difference (see Fig. \ref{Fig:performance_curves}, right). In addition, the higher sensitivity of the ASTRI Mini-Array and in particular of CTA South will allow to enhance the number of VHE detected sources in the TeV band up to a factor of several hundreds \citep{Olmi23} (see Fig. \ref{Fig:performance_curves}, left)

%!!!!!!!!!!!!!!!! 
%!!!!!!!!!!!!!!!!

The ASTRI Mini-Array will have the first three telescopes operative at the beginning at 2024 and the whole array for the first half of 2025. This means that its first results will go parallel or just before the full operation of IceCube-Gen2 (2030s), KM3NeT-ARCA and P-ONE (after 2030) in the next 5-10 years, preparing the ground for a future \gray{}-neutrino combined analysis. 

\begin{figure}[!h]
    \centering
    \includegraphics[width=0.7\textwidth]{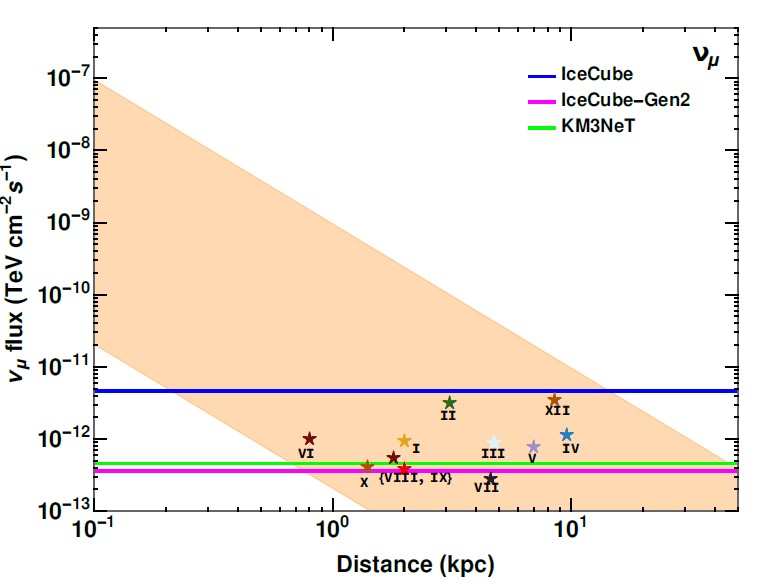}
    \caption{Detection horizon of LHAASO sources (colored stars) for different neutrino detectors (figure and details in \cite{Sarmah23_neutrini}). This is a model dependent estimation but we show it in order to stress the important role of future instruments in the PeVatron contexts. The orange band shows the energy integrated $\nu_{\mu}$ flux as a function of distance for the parameters used in Fig.3 of \cite{Sarmah23_neutrini}.}
    \label{Fig:Sarmah23}
\end{figure}

In \cite{Huang22_neutrini_b}, the authors estimate the observational time it takes for a 5$\sigma$ detection by the so called Planetary Neutrino Monitoring System (PLE$\nu$M) \citep{Schumacher21_neutrini}, a global repository of HE neutrino observations by current and future neutrino telescopes. According their computations (and relative assumption on threshold energy and models), they found that in the next 20 years the PLE$\nu$M coulddiscover neutrino emission from Crab Nebula, LHAASO J1825-1326, LHAASO J1839-0545, LHAASO J1908+0621, LHAASO J2018+3651 and LHAASOJ2226+6057. 
According the models developed in \cite{Sarmah23_neutrini}, IceCube-Gen2 and KM3NeT will be able to detect neutrinos from SNRs (with parameter space based on the two SNRs analysed) with the condition that the neutrino fluxes are above $10^{-3}$ TeV cm$^{-2}$ s$^{-1}$ for $E>10$ TeV and IceCube-Gen2 will be able to detect all the LHAASO source in case they were neutrino emitters (see Fig. \ref{Fig:Sarmah23}). These estimation are strongly dependent on the models used to estimate the parameter space but, however, show that there will be a "neutrino bright" future for the real hadronic PeVatrons.

All the improvement in the technology of future \gray{} and neutrino experiments will have to be accompanied by increasingly detailed theoretical models, able to characterized in a more stringent way the different source types exploiting more detailed experimental data.

%!!!!!!!!!!!!!!!! 
%!!!!!!!!!!!!!!!! 

%%%%%%%%%%%%%%%%%%%%%%%%%%%%%%%%%%%%%%%%%%
\section{Conclusions}
\label{conclusions}
%%%%%%%%%%%%%%%%%%%%%%%%%%%%%%%%%%%%%%%%%%

This is a very exciting moment for the VHE/UHE astrophysics and in particular for the CR origin issue. After just over a century, we have not only a deeper understanding of the physics behind CR acceleration but also a lot of data that are allowing us to correct the path and to know better our Galaxy and their sources. 
We know that Supernova Remnants accelerate the lowest energy component of the Galactic CRs but that we need other types of sources in order to explain CR PeV energies, their spectrum and their composition. Superbubbles and Massive Star Clusters are the major candidates, explaining PeV energies and CR composition, but we need more detailed models in order to understand the acceleration mechanism active in their environment. The standard-leptonic-accelerator PWNe contribute to CR acceleration but likely in a small percentage with respect the other source typologies.

Anyway, most of the LHAASO candidate PeVatrons seem to be associated to PWNe and this can be explained as a statistical bias due to the fact that PWNe are the most numerous sources in our \gray{} bright Galaxy. Only LHAASO J0534+2202 has a firm candidate, the Crab Nebula, instead all the other sources could be associated not only with PWNe but also with SNRs, MSCs and TeV halos. Only a higher sensitivity and spatial resolution multi-wavelength and multi-messengers analysis will be able to make clear most of these confusing association.

LHAASO has demonstrated once again how an improvement in technology leads to an advance in knowledge, especially in the field of astrophysics. 
We can then define an epoch "after LHAASO" ( A.L.) when the term PeVatrons "exploded" in scientific papers \citep{SafiHarb22H} and in this brief review we tried to summarize the huge effort done by the Scientific community during these two last years, based on a strong theory-experimental background before LHAASO (B.L.), to make order in the very large amount of information that we have in our hands. A summary of the main information about every LHAASO source is shown in the Table in Fig.\ref{Fig:Cardillo_table}.

The future instruments for \gray{} and neutrino detection will have a huge responsibility because our comprehension of Galactic Pevatrons will depend on their performances that we hope will be how or better we are estimating during their development. When we will have the first data, there is a large chance to put one more piece into the CR origin puzzle.  At the same time, further measurements of CR spectrum, composition and anisotropy could help us to a better understanding of the physics behind CR acceleration, escape and propagation.
 
We want to conclude with a quote from Felix Aharonian about the origin of CRs : "It is not correct to still speak of mystery because we know a lot of things about them". 
A lot of work has been done with beautiful and outstanding results and a bright future in \gray{} and neutrino is waiting for us.

%%%%%%%%%%%%%%%%%%%%%%%%%%%%%%%%%%%%%%%%%%
\vspace{6pt}

%%%%%%%%%%%%%%%%%%%%%%%%%%%%%%%%%%%%%%%%%%
\authorcontributions{Conceptualization, M.C.; resources, M.C.; writing---original draft preparation, M.C.; writing---review and editing, M.C. and A.G.; All authors have read and agreed to the published version of the manuscript.}

\funding{This research received no external funding}

\acknowledgments{We thank all the referees for their very useful comments and suggestions that have improved our review. We thanks all the speakers of the conference HONEST 2022 (https://indico.desy.de/event/34265/) that gave a very complete and useful picture of the current status of the PeVatron issue.}

\conflictsofinterest{The authors declare no conflict of interest.} 

\abbreviations{Abbreviations}{
The following abbreviations are used in this manuscript:\\

\noindent 
\begin{tabular}{@{}ll}
IACT & Imaging Atmospheric Cherenkov Telescope\\
CR & Cosmic Ray\\
CTA & Cherenkov Telescope Array\\
DSA & Diffusive Shock Acceleration\\
EAS & Extensive Air Shower\\
GMC & Giant Molecular Cloud\\
HE & High Energy\\
IACT & Imaging Atmospheric Cherenkov Telescope\\
IC & Inverse Compton\\
MC & Molecular Cloud\\
MSC & Massive Star Cluster\\
PWN & Pulsar Wind Nebula\\
SNR & Supernova Remnant\\
UHE & Ultra High Energy\\
VHE & Very High Energy\\
YMSC & Young Massive Star Cluster\\
\end{tabular}
}

%%%%%%%%%%%%%%%%%%%%%%%%%%%%%%%%%%%%%%%%%%
\begin{adjustwidth}{-\extralength}{0cm}

\reftitle{References}


%\bibliography{your_external_BibTeX_file}

%%%%%%%%%%%%%%%%%%%%%%%%%
{\scriptsize
\bibliography{Cardillo_PeVatrons.bib}}
%%%%%%%%%%%%%%%%%%%%%%%%

% If authors have biography, please use the format below
%\section*{Short Biography of Authors}
%\bio
%{\raisebox{-0.35cm}{\includegraphics[width=3.5cm,height=5.3cm,clip,keepaspectratio]{Definitions/author1.pdf}}}
%{\textbf{Firstname Lastname} Biography of first author}
%
%\bio
%{\raisebox{-0.35cm}{\includegraphics[width=3.5cm,height=5.3cm,clip,keepaspectratio]{Definitions/author2.jpg}}}
%{\textbf{Firstname Lastname} Biography of second author}

% For the MDPI journals use author-date citation, please follow the formatting guidelines on http://www.mdpi.com/authors/references
% To cite two works by the same author: \citeauthor{ref-journal-1a} (\citeyear{ref-journal-1a}, \citeyear{ref-journal-1b}). This produces: Whittaker (1967, 1975)
% To cite two works by the same author with specific pages: \citeauthor{ref-journal-3a} (\citeyear{ref-journal-3a}, p. 328; \citeyear{ref-journal-3b}, p.475). This produces: Wong (1999, p. 328; 2000, p. 475)

%%%%%%%%%%%%%%%%%%%%%%%%%%%%%%%%%%%%%%%%%%
%% for journal Sci
%\reviewreports{\\
%Reviewer 1 comments and authors’ response\\
%Reviewer 2 comments and authors’ response\\
%Reviewer 3 comments and authors’ response
%}
%%%%%%%%%%%%%%%%%%%%%%%%%%%%%%%%%%%%%%%%%%
\PublishersNote{}
\end{adjustwidth}
\end{document}